\documentclass{article}

\usepackage{amsthm,amsmath,amsfonts,amssymb}
\usepackage{natbib}

\usepackage[latin1]{inputenc}
\usepackage[english]{babel} 
\usepackage{subfigure}
\usepackage{setspace}
\usepackage{tabularx} 
\usepackage{enumerate}
\usepackage{epsfig}
\usepackage{graphics}
\usepackage{graphicx}
\usepackage{color}
\usepackage{url}

\usepackage[short,12hr]{datetime}
\usepackage{fullpage}
\usepackage{srcltx}
\usepackage{flushend}
\usepackage{mathptmx}

\allowdisplaybreaks[4] 
\newcommand{\R}{$\mathbb{R}$}
\newcommand{\tablesize}{\fontsize{7}{11}\selectfont}

\title{Model selection criteria in beta regression with varying dispersion}

\author{Fábio~M.~Bayer%
\thanks{F.~M.~Bayer 
is with the
Departamento de Estat\'istica and LACESM,
Universidade Federal de Santa Maria, RS, Brazil,
E-mail: bayer@ufsm.br}
\and
Francisco~Cribari Neto%
\thanks{F.~Cribari-Neto 
is with the 
Departamento de Estat\'istica, 
Universidade Federal de Pernambuco, PE, Brazil,
E-mail: cribari@de.ufpe.br}
}

\date{}

\begin{document}

\maketitle

\doublespacing

\abstract{
\noindent
We address the issue of model selection in beta regressions with varying dispersion. The model consists of two submodels, namely: for the mean and for the dispersion. Our focus is on the selection of the covariates for each submodel. Our Monte Carlo evidence reveals that the joint selection of covariates for the two submodels is not accurate in finite samples. We introduce two new model selection criteria that explicitly account for varying dispersion and propose a fast two step model selection scheme which is considerably more accurate and is computationally less costly than usual joint model selection. Monte Carlo evidence is presented and discussed. We also present the results of an empirical application.
}


\section{Introduction}

Regression analysis is used for modeling the behavior of a random variable (response, dependent variable) when such a behavior is influenced by other variates (known as regressors, covariates or independent variables). The normal linear regression is the most commonly used regression model. It is not, however, useful for modeling data that assume values in the standard unit interval, $(0,1)$, such as rates and proportions, since it may yield predictions outside the interval. A common practice used to be to transform the data so that they assume values on the real line and then use the transformed response in linear regression analysis. One of the pitfalls of such an approach is that the model parameters can no longer be interpreted in terms of the mean response; their interpretation now involves the mean of the transformed response, which is not of interest. Additionally, rates and proportions are usually asymmetrically distributed and display a particular kind of heteroskedastic behavior. The usual linear regression is thus not appropriate for modeling such data. 

Several practitioners have modeled data that assume values in the standard unit interval 
\citep{Brehm1993, Kieschnick2003, Smithson2006, Zucco2008, Verhaelen2013, Whiteman2013, Hallgren2013}. 
\cite{Ferrari2004} proposed 
a regression model that was specifically tailored for modeling such data: the beta regression model. The underlying assumption is that the response ($y$) is beta-distributed, i.e., it follows the beta law. The beta distribution is quite flexible for modeling rates and proportions since its density can have different shapes depending on the values of its two parameters, mean ($\mu$) and precision ($\phi$)~\citep{Ferrari2004}.

\begin{figure}[!h]
  \centering
  \includegraphics[width=0.6\textwidth]{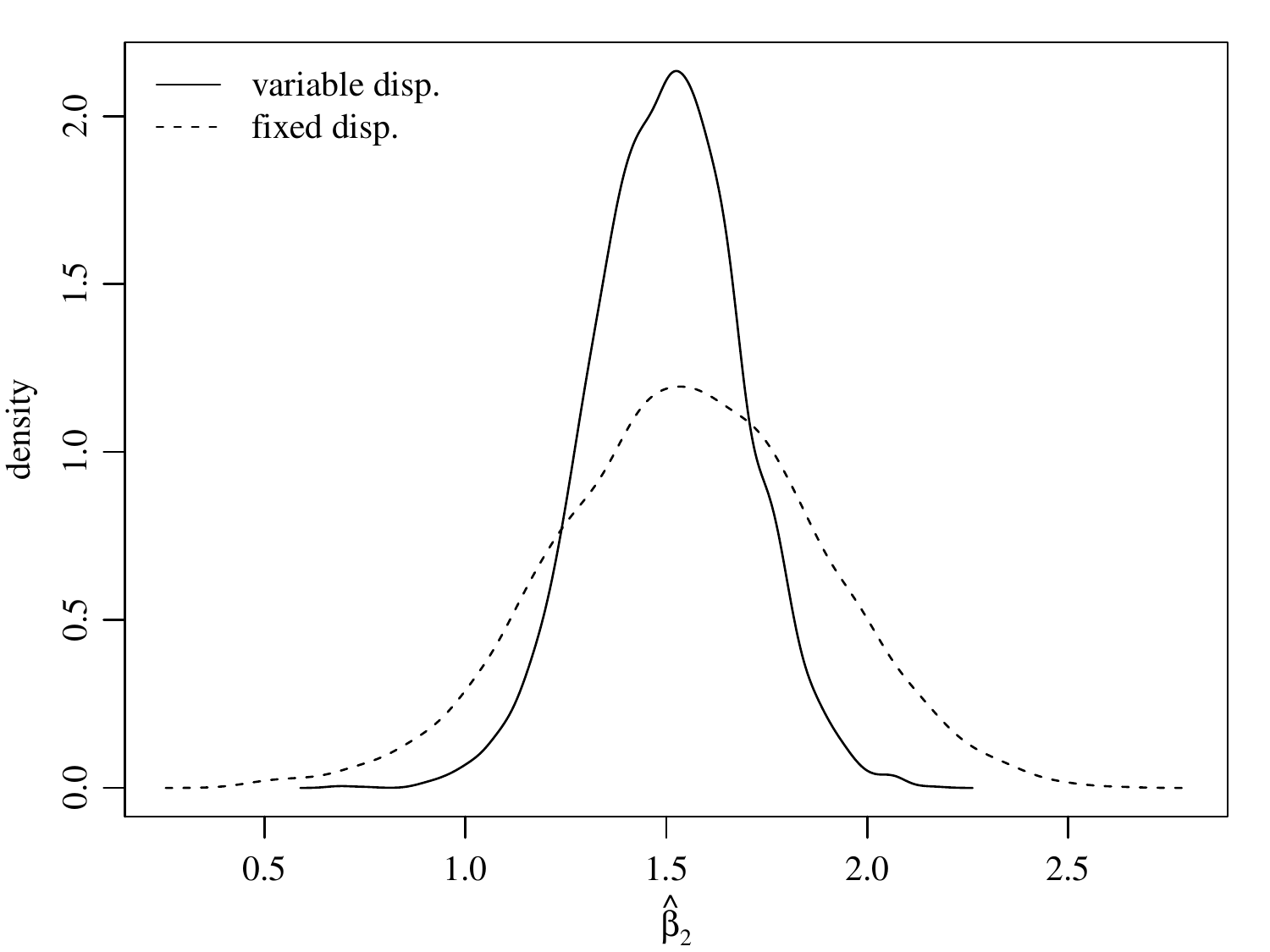}
  \caption{Estimated densities of the slope parameter estimator under varying dispersion, with (continuous line) and without (dashed line) the variations in the precision parameter taken into account.} \label{F:effic}
\end{figure}

In the beta regression model the mean response $\mu$ is related to a linear predictor that includes covariates and unknown regression parameters through a link function in similar fashion to generalized linear models (GLMs) \citep{McCullagh1989}. In its original formulation, the precision parameter $\phi$ was taken to be constant. We note that efficiency loss takes place when the precision parameter is incorrectly taken to be constant. This fact can be seen in Figure~\ref{F:effic}, which presents the estimated densities of maximum likelihood estimates of the slope parameter ($\beta_2=1.5$) in a single covariate model under varying dispersion. The density estimates were constructed from a Monte Carlo simulation with five thousand replications. 
The data generating process is $\rm{logit}(\mu_t)=\beta_1 + \beta_2 x_{t}$  
with precision given by $\log(\phi_t)=\gamma_1 + \gamma_2 x_{t}$.
The two densities correspond to the situations in which dispersion was incorrectly taken to be fixed (`fixed disp.') and properly modeled (`variable disp.'). Notice that the variance is considerably larger when the dispersion is not modeled. 
Additionally, disperion modelling may be of direct interest since it allows the statistician to identify the sources of
data variability \citep{Smyth1999, Wu2012}.

Figure \ref{F:effic} shows that efficient parameter estimation in regression model depends on the correct modeling of the dispersion. In the class of GLMs, \cite{Smyth1989}, \cite{Nelder1991} and \cite{Smyth1999} define a joint generalized linear model, which allows the joint modeling of the response mean and variance. In this perspective, 
\cite{Smithson2006}, \cite{Ospina2007} and \cite{Simas2010} suggest the beta regression model with varying dispersion.
This model 
can be seen as a natural extension of the model introduced by \cite{Ferrari2004}. The precision parameter now relates to a set of covariates and parameters through a link function. The model thus includes two submodels: one for the mean response and another one for the precision.   

Our goal in this paper is threefold. 
First, we present several model selection criteria and propose two new criteria that explicitly account for varying dispersion. 
Second, we perform a Monte Carlo simulation study to compare the finite sample performances of the traditional and new model selection criteria. 
The numerical evidence shows that the joint selection of the regressors in both submodels can be quite unreliable. 
Thirdly, we then propose a fast two step procedure that works 
better in finite samples and is computationally less costly than the joint covariates selection. 
Our proposal is to perform model selection for the mean submodel taking the precision to be constant and only in a second step to carry out model 
selection for the precision submodel. 
The likelihood of finding the correct model is 
increased and the computational cost is greatly reduced when the proposed two step procedure is used.

The paper unfold as follows. In the next section we present the varying dispersion beta regression model. In Section~\ref{s:MS} we describe different model selection strategies, 
including our proposed fast two step model selection scheme.
Section~\ref{s:Numerical} presents numerical evidence from Monte Carlo simulations 
and a guideline for choose model selection criteria.
The evidence favours the model selection scheme proposed in this paper. Section~\ref{S:application} contains an empirical application. Finally, concluding remarks are offered in Section~\ref{S:conclusions}.

\section{The model and parameter estimation}\label{S:model_beta_var}

Let $y$ be a random variable that follows the beta law with parameters $\mu$ and $\phi$. In what follows we use an alternative parametrization, namely: $\sigma^2=1/(1+\phi)$. Both parameters ($\mu$ and $\sigma$) assume values in the standard unit interval.\footnote{Notice that $\sigma$ is a dispersion, not precision, parameter.} Thus, 
\begin{align}
& {\rm E}(y)=\mu, \nonumber \\
& {\rm var}(y)=V(\mu)\sigma^2.\label{E:vary}
\end{align}
Additionally, the density of $y$ can be written as
\begin{equation}\label{E:densidade2a}
f(y;\mu,\sigma)=
\frac{\Gamma\left(\frac{1-\sigma^2}{\sigma^2}\right)}{\Gamma\left(\mu\left(\frac{1-\sigma^2}{\sigma^2}\right)\right)
\Gamma\left(\left(1-\mu\right)\left(\frac{1-\sigma^2}{\sigma^2}\right)\right)}
y^{\mu\left(\frac{1-\sigma^2}{\sigma^2}\right)-1}(1-y)^{(1-\mu)\left(\frac{1-\sigma^2}{\sigma^2}\right)-1}, \quad 0<y<1,
\end{equation}
where $0<\mu<1$ and $0<\sigma<1$.

Let $y_1,\ldots,y_n$ be independent random variables, each $y_t$, $t=1,\ldots,n$, having density  \eqref{E:densidade2a} with mean $\mu_t$ and dispersion $\sigma_t$. The beta regression model with varying dispersion can be written as 
\begin{align}
& g(\mu_t)=\sum_{i=1}^{r}x_{ti}\beta_i=\eta_t, \nonumber \\
& h(\sigma_t)=\sum_{i=1}^{s}z_{ti}\gamma_i=\nu_t, \label{E:h_sigma}
\end{align}
where $\beta = (\beta_1,\ldots,\beta_r)^{\top}$ and $\gamma = (\gamma_1,\ldots,\gamma_s)^{\top}$ are unknown parameters. Additionally, $x_{t1},\ldots,x_{tr}$ and $z_{t1},\ldots,z_{ts}$ are independent variables ($r+s=k<n$). When intercepts are included in both submodels we have that $x_{t1}=z_{t1} = 1$,  $t=1,\ldots,n$. Finally, $g(\cdot)$ and $h(\cdot)$ are the strictly monotonic and twice differentiable link functions that map $(0,1)$ into \R. Commonly used link functions are logit, probit, log-log, complement log-log and Cauchy. Note that under the current parametrization the same link functions that are used in the mean submodel can also be used in the dispersion submodel. 
This is the parametrization also used by \cite{CribariSouza2012}. 

Estimation of $\beta$ and $\gamma$ can be carried out by maximum likelihood. The log-likelihood function is 
\begin{equation*}
\ell ({\beta},{\gamma})=\sum^n_{t=1}\ell_t(\mu_t,\sigma_t),
\end{equation*}
where
\begin{align*}
\ell_t(\mu_t,\sigma_t) &= \log \Gamma{\left( \frac{1-\sigma^2_t}{\sigma^2_t} \right)} \! - \! \log \Gamma{\left(\! \mu_t \left(\frac{1-\sigma^2_t}{\sigma^2_t}\right) \!\!\right)} \! - \! \log \Gamma{\left(\! (1-\mu_t) \!  \left(\frac{1-\sigma^2_t}{\sigma^2_t} \right) \!\! \right)}\\
& + \left[\mu_t \left(\frac{1-\sigma^2_t}{\sigma^2_t} \right)-1 \right] \log y_t + \left[(1-\mu_t)\left(\frac{1-\sigma^2_t}{\sigma^2_t}\right)-1 \right]\log(1-y_t). 
\end{align*}
Details on the score function $U(\beta, \gamma)$, 
Fisher's information matrix $K({\beta}, {\gamma})$,  
and large sample inferences can be found in \cite{CribariSouza2012}.

It is noteworthy that it is possible to test whether dispersion is constant, i.e., test 
the null hypothesis
$$
\mathcal{H}_0 : \sigma_1 = \sigma_2 = \cdots = \sigma_n = \sigma,
$$
or, equivalently,
$$
\mathcal{H}_0 : \gamma_i = 0, \quad i=2,\ldots,s,
$$
for the model given in \eqref{E:h_sigma} with $z_{t1}=1$, $t=1,\ldots,n$. The score statistic is
\begin{equation*}
S=\widetilde{U}^{\top}_{(s-1)\gamma}\widetilde{K}^{-1}_{(s-1)(s-1)}\widetilde{U}_{(s-1)\gamma},
\end{equation*}
where $\widetilde{U}^{\top}_{(s-1)\gamma}$ is the vector with the $s-1$ final elements of the score function for ${\gamma}$ under $\mathcal{H}_0$ and  $\widetilde{K}^{-1}_{(s-1)(s-1)}$ is the $(s-1)\times(s-1)$ matrix that contains the last $s-1$ rows and the last $s-1$ columns of the inverse of Fisher's information matrix evaluated at the restricted maximum likelihood estimator. 
Under the usual regularity conditions and under $\mathcal{H}_0$, $S$ converges in distribution to $\chi^2_{(s-1)}$. The null hypothesis is thus rejected if $S > \chi^2_{1-\alpha, s-1}$, where $\chi^2_{1-\alpha, s-1}$ is the $1-\alpha$ $\chi^2_{s-1}$ upper quantile, $\alpha$ being the test nominal level.

\section{Model selection criteria}\label{s:MS}

Model selection in regression analysis is of paramount importance. Three important decisions are typically made: (i) assuming a response distribution; (ii) selection the link functions to be used and (iii) choosing which regressions are to included in the linear predictor(s).
The beta distribution is typically adequate for modeling continuous random variables that assume values in $(0,1)$.  
We also note that the correct specification of the link function(s) can be assessed using the misspecification test proposed in \cite{Pereira2013}. 
It remains to address the decision outlined in item (iii), i.e., covariates selection. We shall do so in what follows.
 
Several model selection criteria were proposed for the linear regression model. The first widely used criterion was the adjusted $R^2$. It penalizes the coefficient of determination ($R^2$) whenever more regressors are added to the model. Other commonly used model selection criteria are the AIC \citep{Akaike1973, Akaike1974}, Mallows's $Cp$ \citep{Mallows1973}, the BIC \citep{Akaike1978} or SIC \citep{Schwarz1978}, and the HQ \citep{Hannan1979}. Some of these criteria are also used in the class of generalized linear models. \cite{Hu2008} introduced a class of consistent criteria based on a modification of the $R^2$ statistic. A good reference on model selection in linear regression is \cite{McQuarrie1998}. We note that a selection criterion is consistent when it identifies the finite dimension correct model asymptotically with probability one \citep{McQuarrie1998} provided that the true model is among the candidate models. 

To the best of our knowledge there 
are no results in the literature on model selection 
criteria in the class of beta regression with varying dispersion.
We note, however, that several of the usual model selection criteria can be used in such a class of models. 
In what follows we shall review them. 
Nevertheless, in a different sense, an alternative to our approach is presented in \cite{Shou2013}. 
The authors compare two measures for simultaneously evaluating 
the relative importance of predictors in location and dispersion submodels in beta regression, 
but without focusing on model selection.

\subsection{Usual model selection criteria}

At the outset, we introduce a penalized version of the beta regression pseudo-$R^2$ used in \cite{Ferrari2004}. This pseudo-$R^2$ is defined as the square of the sample coefficient of correlation between $g({y})$ and $\widehat{\eta}=X\widehat\beta$, where $\widehat\beta$ denotes the maximum likelihood estimator of $\beta$ and $X$ is the $n\times r$ matrix of covariates used in the mean submodel. We penalize their pseudo-$R^2$ so that it includes a penalty term that takes into account the model dimension. The penalized pseudo-$R^2$ criterion is given by  
\begin{equation*}
\bar{R}^2_{FC}=1-(1-R^2_{FC})\frac{(n-1)}{(n-k)},
\end{equation*}
where $R^2_{FC}$ is the pseudo-$R^2$ proposed in \cite{Ferrari2004} and $k=r+s$ is the number of estimated parameters of the model.

Let $L_{\rm fit}$ and $L_{\rm null}$ denote, respectively, the maximized log-likelihood functions of the beta regression model and of the model without covariates (with only the intercepts included in the two submodels). Following \cite{Nagelkerke1991} and \cite{Long1997}, a measure of goodness-of-fit can be written as
\begin{equation*}
R^2_{LR} = 1-\left(\frac{L_{\rm null}}{L_{\rm fit}}\right)^{2/n}.
\end{equation*}
We penalize this quantity  
and define the following model selection criterion:
\begin{equation}\label{E:R2ML}
\bar{R}^2_{LR}=1-(1-R^2_{LR})\frac{(n-1)}{(n-k)}.
\end{equation}

The proposal of an $R^2$ measure for GLMs can be found in \cite{Hu2008}. Using this measure of goodness-of-fit, they proposed a model selection criterion, which is given by 
\begin{equation*}
\bar{R}^2_{HS}=1-\frac{n-1}{n-\lambda_n k}\frac{\sum_{t=1}^n (y_t-\widehat{\mu}_t)^2}{\sum_{t=1}^n (y_t-\bar{y})^2},
\end{equation*} 
where $\bar{y}=\frac{1}{n}\sum_{t=1}^n y_t$ and $\widehat{\mu}_t=g^{-1}(\widehat{\eta}_t)$. If $\lambda_n = 1$, then $\bar{R}^2_{HS}$ reduces to the modified $R^2$ given in \cite{Mittlbock2002}. Additionally, if $\lambda_n =o(n)$ and $\lambda_n \rightarrow \infty$ when $n\rightarrow \infty$, then the criterion is consistent \citep{Hu2008}. The authors recommend using $\lambda_n = 1$, $\lambda_n = \log{(n)}$ or $\lambda_n = \sqrt{n}$.

The model selection criteria presented so far are based on measures of goodness-of-fit. The best model is thus that which maximizes the criterion. An alternative is to define criteria that must be minimized (rather than maximized) when searching for the model that best fits the data. A well known example is the Akaike information criterion (AIC) \citep{Akaike1973, Akaike1974}:  
\begin{equation*}
\textrm{AIC}=-2\ell(\widehat\beta,\widehat\gamma)+2k,
\end{equation*}
where $\widehat\beta$ and $\widehat\gamma$ are the maximum likelihood estimators of the $\beta$ and $\gamma$, respectively.

Assume that the true model has infinite dimension and that the set of candidate models does not contain the true model. According to \cite{Shibata1980}, a model selection criterion is said to be asymptotically efficient if, in large samples, it selects the model that minimizes the mean squared difference between $\mu$ and $\widehat\mu=g^{-1}(\widehat\eta)$. The AIC, for instance, is asymptotically efficient. Recall, nonetheless, that it was derived as an asymptotically unbiased estimator of the Kullback-Leibler distance \citep{Kullback1951} between the true model and the candidate (estimated) model \citep{Akaike1973,Bengtsson2006}. It is thus based on a large sample approximation and may not deliver accurate model selection when the sample size is small.  \cite{Sugiura1978} introduces an unbiased estimator for the Kullback-Leibler distance in linear regressions: the AICc. \cite{Hurvich1989} generalize AICc to nonlinear regressions and autoregressive models, showing that it is asymptotically equivalent to the AIC but delivers more reliable model selection in finite samples. The AICc is defined as
\begin{equation*}
\textrm{AICc}=-2\ell(\widehat\beta,\widehat\gamma)+\frac{2nk}{n-k-1}.
\end{equation*}

Using a Bayesian approach, \cite{Akaike1978} and \cite{Schwarz1978} introduced a consistent model selection criterion for the linear regression model. The Schwarz information criterion (SIC), also known as the Bayesian information criterion (BIC), is given by
\begin{equation*}
\textrm{SIC}=-2\ell(\widehat\beta,\widehat\gamma)+k\log(n).
\end{equation*}

\cite{McQuarrie1999} derived a version of the SIC that includes a small sample correction, namely: the SICc. Like the SIC, the SICc is consistent. It is given by
\begin{equation*}
\textrm{SICc}=-2\ell(\widehat\beta,\widehat\gamma)+\frac{nk\log(n)}{n-k-1}.
\end{equation*}

Another consistent criterion is the HQ criterion, which was proposed by \cite{Hannan1979} for autoregressive model selection:
\begin{equation*}
\textrm{HQ}=-2\ell(\widehat\beta,\widehat\gamma)+2k\log(\log(n)).
\end{equation*}
A variant of the HQ that incorporates a finite sample correction was proposed by \cite{McQuarrie1998} and is given by 
\begin{equation*}
\textrm{HQc}=-2\ell(\widehat\beta,\widehat\gamma)+\frac{2nk \log(\log(n))}{n-k-1}.
\end{equation*}

\subsection{Model selection criteria under varying dispersion} \label{SS:crit_var_disp}

The usual model selection criteria do not explicitly account for varying dispersion. They typically use the distance between $y$ and $\widehat\mu$ as a goodness-of-fit measure to be penalized by the inclusion of extra covariates in the model. We shall now propose two new model selection criteria that take into account the information that precision is not constant and is modeled alongside with the mean.

The inclusion of covariates in the mean and dispersion submodels may impact the goodness-of-fit in different ways. In order to account for that, 
we introduce, based on $R^2_{LR}$, the weighted $\bar{R}^2_{LR}$, given by 
\begin{equation*}
\bar{R}^2_{LRw}=1-(1-R^2_{LR})\left(\frac{n-1}{n-(1+\alpha)r - (1-\alpha)s}\right)^{\delta},
\end{equation*}
where $0 \leq \alpha \leq1$ and $\delta > 0$. We note that when $\alpha=0$ and $\delta=1$ the above criterion reduces to $\bar{R}^2_{LR}$ given in \eqref{E:R2ML}. The latter is thus a particular case of the former. Our Monte Carlo evidence in Section \ref{s:Numerical} will sheds some light on the choice of values for $\alpha$ and $\delta$.

A second model selection criterion we introduce for covariates selection under varying dispersion is based on a convex combination of the mean and dispersion goodness-of-fit measures both penalized by the respective number of regressors.  
From \eqref{E:vary}, 
$\mathrm{var}(y_t)=\sigma_t^2\mu_t(1-\mu_t)$, i.e.,  
$\sigma_t^2 =\mathrm{var}(y_t)/\mu_t(1-\mu_t).$
We note that $\mathrm{var}(y_t)$ can be approximated by $(y_t-\widehat\mu_t)^2$, 
and define $\sigma^{\ast}_t=\sqrt{\frac{(y_t-\widehat{\mu}_t)^2}{\widehat{\mu}_t(1-\widehat{\mu}_t)}}$. 
Based on $\bar{R}^2_{HS}$~\citep{Hu2008}, 
we then propose the following model selection criterion for varying dispersion models: 
\begin{equation*}
\bar{R}^2_{D} \! = \! \alpha\left[1 \! - \! \frac{n-1}{n-\lambda_n r}\frac{\sum_{t=1}^n (y_t-\widehat{\mu}_t)^2}{\sum_{t=1}^n (y_t-\bar{y})^2}\right] + (1-\alpha) \! \left[1 \! - \! \frac{n-1}{n-\delta_n s}\frac{\sum_{t=1}^n (\sigma^{\ast}_t-\widehat{\sigma}_t)^2}{\sum_{t=1}^n (\sigma^{\ast}_t-\bar{\sigma}^{\ast})^2}\right]\!,
\end{equation*}
where $\bar{\sigma}^{\ast}=(1/n)\sum_{t=1}^{n}\sigma^{\ast}_t$, $0\leq\alpha\leq1$ and $\delta_n$, as well as $\lambda_n$ for $\bar{R}^2_{HS}$, is a function of $n$, such as, for example, $\delta_n = 1$, $\delta_n = \log{(n)}$ and $\delta_n = \sqrt{n}$.

\subsection{Proposed fast two step model selection scheme}\label{ss:two-step}

The criteria presented so far are typically used for the joint selection of the mean and dispersion regressors. 
However, the numerical results presented in Section~\ref{s:Numerical} 
show that such a joint selection may be quite inaccurate in finite samples.
Furthermore, it can computationally unfeasible even when the number of candidate covariates is moderate. 
In what follows we propose a two-step model selection scheme, which is more accurate and more computationally efficient that joint 
covariates selection.

In order to reduce the varying dispersion beta regression model selection computational cost and motivated by the fact that the 
criteria that perform well for covariates selection in the mean submodel may not perform equally well when the focus lies in 
selection regressors for the dispersion submodel, we introduce a model selection strategy that consists of two steps, which are 
performed sequentially. The scheme can be outlined as follows:  
\begin{enumerate}
\item[(1)] assuming constant dispersion, select regressors for the mean submodel;  
\item[(2)] assuming that the mean submodel selected in Step (1) is adequate, use a model selection criterion to select regressors for the dispersion submodel.
\end{enumerate} 

The proposed scheme has the advantage being computationally less costly than the joint model selection. 
Suppose there are $m$ candidate regressors for the mean and dispersion submodels. 
Joint model selection of the two submodels entails the estimation $(2^m+1)^2$ different models. 
The proposed fast scheme requires estimation of only $2\times(2^m+1)$ models.  The ratio of these figures is
\begin{align*}
\frac{(2^m+1)^2}{2(2^m+1)}=\frac{(2^m+1)}{2}\approx2^{m-1}.
\end{align*}
The proposed method is thus approximately $2^{m-1}$ times less computationally intensive than the usual approach. 

For instance, consider $m=10$ (ten candidate covariates for the two submodels). Joint model selection of the mean and dispersion submodels requires one to estimate $(2^{10}+1)^2=1050625$ beta regressions whereas our model selection strategy only entails the estimation of $2\times(2^{10}+1)=2050$ models.
The computational efficiency factor thus equals $512.5\approx 2^{10-1}=512$. 
Suppose that it takes one second to fit a beta regression model. The joint scheme would then run for 12 days whereas our scheme would 
only take 34 minutes. 

In the Section \ref{s:Numerical}, 
we present Monte Carlo evidence on the proposed model selection scheme. We used different combinations of criteria in Steps (1) and (2), based on some numerical evidences.

\section{Numerical evaluation}\label{s:Numerical}

We shall now report the results of a set of Monte Carlo simulations that were carried out to assess the relative merits of the different model selection criteria in varying dispersion beta regressions. All simulations were performed using the statistical computing environment \textsc{R} (version 2.9) \citep{R2009}. Parameter estimation was performed using the {\tt GAMLSS} package \citep{Stasinopoulos2007}. 
 An implementation of our two-step scheme in {\tt R} language is available at \url{http://www.ufsm.br/bayer/auto-beta-reg.zip}. 
This file contains model selection computer code and also the dataset used in the empirical application presented in Section~\ref{S:application}. 

The beta regression model used as data generating process is 
\begin{align}
g(\mu_t) &= \beta_1+ x_{t2} \beta_2 + x_{t3} \beta_3 + x_{t4} \beta_4 + x_{t5} \beta_5, \label{E:mod_mu}\\
h(\sigma_t)&= \gamma_1+ z_{t2} \gamma_2 + z_{t3} \gamma_3 + z_{t4} \gamma_4 + z_{t5} \gamma_5,  \label{E:mod_sigma}
\end{align}
$t=1,\ldots,n$, where \eqref{E:mod_mu} is the mean submodel, \eqref{E:mod_sigma} is the dispersion submodel and $x_{ti}=z_{ti}$, $i=2,\ldots,5$ and $\forall t$. We used different values for the parameter vector ${\theta}=(\beta_1,\beta_2,\beta_3,\beta_4,\beta_5,\gamma_1, \gamma_2, \gamma_3, \gamma_4, \gamma_5)^\top$ and also four different sample sizes: $n=25, 50, 100, 200$. The parameter values are presented in Table~\ref{T:models}. The number of Monte Carlo replications was 5,000. All covariate values were obtained as random draws from the standard uniform distribution $\mathcal{U} (0,1)$ and were kept fixed throughout the experiment.

\begin{table}[ht]
\begin{center}
\caption{Parameter values using in the data generating process.}\label{T:models}
{
\tablesize
\begin{tabular}{lrrrrrrrrrr}
  \hline
 Models & $\beta_1$ & $\beta_2$ & $\beta_3$ & $\beta_4$ & $\beta_5$ & $\gamma_1$ & $\gamma_2$ & $\gamma_3$ & $\gamma_4$ & $\gamma_5$ \\
  \hline
 Model 1 &   $1.5$ & $-1$ & $-1$ & $0$ & $0$ & $-1$ & $-1$ & $-1$ & $0$ & $0$\\
 Model 2 &   $-1.5$ & $1$ & $1$ & $0$ & $0$ & $-1$ & $-1.25$ & $-1/2$ & $-1/4$ & $0$ \\
 Model 3 &   $1$ & $-3/4$ & $-1/4$ & $0$ & $0$ & $-1$ & $-1$ & $-1$ & $0$ & $0$\\
 Model 4 &   $-1$ & $3/4$ & $1/4$ & $0$ & $0$ & $-1$ & $-1.25$ & $-1/2$ & $-1/4$ & $0$ \\
  \hline
\end{tabular}}
\end{center}
\end{table}

Model 1 is easily identifiable since all slopes have the same value. In Model 2, the mean submodel is easily identifiable and the dispersion submodel is weakly identifiable. 
Weak identifiability happens when $\gamma_i$ approaches zero as $i$ grows. Here, the covariates influence the mean response with different intensities. In Model 3, the mean submodel is weakly identifiable and the dispersion submodel is easily identifiable. Finally, both submodels of Model 4 are weakly identifiable. 
For details on such a model identifiability concept, see \cite{McQuarrie1998}, \cite{Caby2000} and \cite{Frazer2009}. 
We emphasize that it differs from the usual concept of model identifiability, which relates to the uniqueness of the model for a given set of parameter values \citep{Paulino1994, Rothenberg1971}.
 
In each Monte Carlo simulation, we generated the responses from the beta distribution with parameters $\mu_t$ and $\sigma_t$, which are given in \eqref{E:mod_mu} and \eqref{E:mod_sigma}, respectively. We used the logit link in both submodels. The data generating process used in our simulations is 
\begin{equation*}
\mu_t=\frac{\exp(\beta_1 + \sum_{i=2}^5{x_{ti} \beta_i})}
{1 +\exp(\beta_1 + \sum_{i=2}^5{x_{ti} \beta_i})}, \;\; 
\sigma_t=\frac{\exp(\gamma_1 + \sum_{i=2}^5{z_{ti} \gamma_i})}
{1 +\exp(\gamma_1 + \sum_{i=2}^5{z_{ti} \gamma_i})}.
\end{equation*}

The set of candidate models includes all models with intercepts that are particular cases of the above model.  
Since there are four regressors in the mean submodel its total number of candidate models is $(2^4 +1) = 17$; likewise for the dispersion submodel. If we take the two submodels together, then there are $ 17 \times 17 =  289$ candidate models that need to be considered. 
For performance evaluation of the model selection criteria, 
we consider the methodology used in \cite{Hannan1979, Hurvich1989, Shao1996, McQuarrie1997, McQuarrie1998, Pan1999, Shi2002, Shang2008, Hu2008, Liang2008}. 
We compute and report the frequency with which each criterion was able to identify the true model. That is, we report the percentages of the 5,000 Monte Carlo replications in which the criteria selected the true model. 

The following approaches were considered in our numerical evaluation: 
\begin{enumerate}
\item we used the model selection criteria to jointly select the regressors of both submodels (i.e., mean and dispersion submodels); 
\item the mean submodel was correctly specified and we focused on selecting the covariates that should be included in the dispersion submodel; 
\item the dispersion submodel was correctly specified and the model selection criteria were used to select  independent variables for the mean submodel; 
\item we assumed that the dispersion parameter was constant and only selected regressors for the mean submodel.
\item based on the first four approaches results, we propose a two-step model selection strategy.
\end{enumerate}
The frequencies (\%) of correct model selection for the five approaches listed above are given in Tables \ref{T:P4}, \ref{T:P3}, \ref{T:P2}, \ref{T:P5} and \ref{T:steps},
respectively. All entries are percentages and the figure corresponding to the best performer is displayed in boldface. 

The criterion $\bar{R}^2_{HS}$ was used with  $\lambda_n = 1$, $\lambda_n = \log{(n)}$ and $\lambda_n = \sqrt{n}$. We shall only present the results obtained using $\lambda_n = \log{(n)}$ since this choice led to the most accurate model selections. Additionally, the criteria discussed in Section~\ref{SS:crit_var_disp} were implemented as follows: 
\begin{enumerate}[]
\item $\bar{R}^2_{D1}$: uses $\alpha=0.4$, $\lambda_n=\log(n)$ and $\delta_n=\log(n)$;
\item $\bar{R}^2_{D2}$: uses $\alpha=0.6$, $\lambda_n=\log(n)$ and $\delta_n=\log(n)$;
\item $\bar{R}^2_{D3}$: uses $\alpha=0.6$, $\lambda_n=\log(n)$ and $\delta_n=1$;
\item $\bar{R}^2_{D4}$: uses $\alpha=0.5$, $\lambda_n=\log(n)$ and $\delta_n=1$;
\item $\bar{R}^2_{LR w1}$: uses $\alpha= 0$ and $\delta=3$;
\item $\bar{R}^2_{LR w2}$: uses $\alpha= 0$ and $\delta=2$;
\item $\bar{R}^2_{LR w3}$: uses $\alpha= 0$ and $\delta=1.5$;
\item $\bar{R}^2_{LR w4}$: uses $\alpha= 0.4$ and $\delta=1$;
\item $\bar{R}^2_{LR w5}$: uses $\alpha= 0.4$ and $\delta=2$.
\end{enumerate}

The choice of values for $\alpha$, $\delta$, $\lambda_n$ and $\delta_n$ used in the variations of  $\bar{R}^2_{D}$ and $\bar{R}^2_{LR w}$ was based on numerical results obtained from pilot simulations. Notice that when the value of $\alpha$ is greater than 0.5 in $\bar{R}^2_{D}$ we give more weight to the dispersion submodel fit. Likewise, values of $\delta$ greater than one make $\bar{R}^2_{LR w}$ penalize more heavily the inclusion of new covariates in the model, the inclusion of new regressors in the mean submodel being more heavily penalized when $\alpha > 0$.

\begin{table}[t]
\tablesize																
{
\caption{Frequencies (\%) of correct joint model selection (jointly selecting regressors for both submodels).} \label{T:P4}	
\begin{center}																
\begin{tabular}{l|rrrr|rrrr|rrrr|rrrr}
\hline																	
 & \multicolumn{4}{c|}{Model $1$}& \multicolumn{4}{c|}{Model $2$}& \multicolumn{4}{c|}{Model $3$}& \multicolumn{4}{c}{Model $4$}\\
\hline	
$n$	& $	25	$ & $	50	$ & $	100	$ & $	200	$ & $	25	$ & $	50	$ & $	100	$ & $	200	$ & $	25	$ & $	50	$ & $	100	$ & $	200	$ & $	25	$ & $	50	$ & $	100	$ & $	200	$ \\
\hline
AIC	 & $	4.9	$ & $	24.0	$ & $	42.9	$ & $	49.3	$ & $	1.2	$ & $	4.4	$ & $	11.3	$ & $	24.2	$ & $	2.3	$ & $	14.5	$ & $	39.6	$ & $	49.1	$ & $	0.7	$ & $	2.9	$ & $	10.0	$ & $	23.6	$ \\
AICc	 & $	3.9	$ & $	27.2	$ & $	49.0	$ & $	53.0	$ & $	0.6	$ & $	3.2	$ & $	10.8	$ & $	23.9	$ & $	1.4	$ & $	\mathbf{15.4}	$ & $	43.8	$ & $	52.5	$ & $	0.3	$ & $	2.0	$ & $	9.5	$ & $	23.6	$ \\
SIC	 & $	3.8	$ & $	24.6	$ & $	\mathbf{64.4}	$ & $	89.5	$ & $	0.7	$ & $	1.4	$ & $	3.7	$ & $	10.4	$ & $	1.6	$ & $	10.2	$ & $	47.8	$ & $	88.0	$ & $	0.4	$ & $	0.5	$ & $	2.8	$ & $	8.9	$ \\
SICc	 & $	1.4	$ & $	20.0	$ & $	63.5	$ & $	\mathbf{91.3}	$ & $	0.2	$ & $	0.6	$ & $	2.2	$ & $	8.8	$ & $	0.3	$ & $	6.2	$ & $	44.4	$ & $	\mathbf{89.4}	$ & $	0.1	$ & $	0.1	$ & $	1.8	$ & $	7.5	$ \\
HQ	 & $	4.7	$ & $	26.3	$ & $	58.7	$ & $	73.5	$ & $	1.1	$ & $	3.0	$ & $	8.0	$ & $	19.8	$ & $	2.2	$ & $	14.5	$ & $	49.4	$ & $	72.4	$ & $	0.6	$ & $	1.9	$ & $	6.9	$ & $	18.6	$ \\
HQc	 & $	3.2	$ & $	26.5	$ & $	62.6	$ & $	76.7	$ & $	0.5	$ & $	1.5	$ & $	6.3	$ & $	18.5	$ & $	0.9	$ & $	12.4	$ & $	\mathbf{51.6}	$ & $	75.5	$ & $	0.2	$ & $	0.8	$ & $	5.3	$ & $	17.3	$ \\
$ \bar{R}^2_{FC}	$ & $	0.0	$ & $	0.0	$ & $	0.0	$ & $	0.0	$ & $	0.0	$ & $	0.0	$ & $	0.0	$ & $	0.0	$ & $	0.0	$ & $	0.0	$ & $	0.0	$ & $	0.0	$ & $	0.0	$ & $	0.0	$ & $	0.0	$ & $five	0.0	$ \\
$\bar{R}^2_{LR}	$ & $	4.2	$ & $	14.7	$ & $	19.8	$ & $	22.0	$ & $	1.5	$ & $	5.4	$ & $	11.5	$ & $	17.7	$ & $	2.2	$ & $	10.5	$ & $	19.8	$ & $	21.8	$ & $	0.9	$ & $	4.2	$ & $	10.9	$ & $	18.8	$ \\
$\bar{R}^2_{HS}	$ & $	0.0	$ & $	0.0	$ & $	0.0	$ & $	0.0	$ & $	0.0	$ & $	0.0	$ & $	0.0	$ & $	0.0	$ & $	0.0	$ & $	0.0	$ & $	0.0	$ & $	0.0	$ & $	0.0	$ & $	0.0	$ & $	0.0	$ & $	0.0	$ \\
$\bar{R}^2_{D1}	$ & $	4.4	$ & $	15.4	$ & $	36.5	$ & $	58.7	$ & $	0.3	$ & $	0.7	$ & $	2.2	$ & $	6.8	$ & $	1.9	$ & $	7.4	$ & $	29.7	$ & $	60.8	$ & $	0.2	$ & $	0.5	$ & $	1.9	$ & $	7.7	$ \\
$\bar{R}^2_{D2}	$ & $	6.0	$ & $	17.9	$ & $	46.4	$ & $	73.9	$ & $	0.3	$ & $	0.8	$ & $	2.7	$ & $	8.7	$ & $	1.5	$ & $	6.3	$ & $	34.0	$ & $	74.0	$ & $	0.1	$ & $	0.4	$ & $	2.0	$ & $	8.7	$ \\
$\bar{R}^2_{D3}	$ & $	\mathbf{14.7}	$ & $	23.6	$ & $	35.1	$ & $	40.0	$ & $	\mathbf{4.9}	$ & $	\mathbf{9.3}	$ & $	16.7	$ & $	27.3	$ & $	3.7	$ & $	8.0	$ & $	26.3	$ & $	40.0	$ & $	1.1	$ & $	3.7	$ & $	12.2	$ & $	28.4	$ \\
$\bar{R}^2_{D4}	$ & $	12.8	$ & $	22.1	$ & $	32.3	$ & $	37.6	$ & $	4.6	$ & $	8.8	$ & $	15.2	$ & $	25.4	$ & $	\mathbf{4.0}	$ & $	8.0	$ & $	24.7	$ & $	37.6	$ & $	1.3	$ & $	4.0	$ & $	11.7	$ & $	27.2	$ \\
$\bar{R}^2_{LRw1}	$ & $	3.3	$ & $	26.5	$ & $	60.3	$ & $	70.5	$ & $	0.6	$ & $	1.9	$ & $	7.6	$ & $	21.0	$ & $	1.0	$ & $	12.8	$ & $	50.4	$ & $	69.1	$ & $	0.2	$ & $	0.9	$ & $	6.3	$ & $	20.0	$ \\
$\bar{R}^2_{LRw2}	$ & $	4.9	$ & $	25.9	$ & $	45.5	$ & $	51.0	$ & $	0.9	$ & $	3.8	$ & $	11.1	$ & $	24.0	$ & $	2.2	$ & $	15.2	$ & $	41.7	$ & $	50.6	$ & $	0.5	$ & $	2.6	$ & $	9.8	$ & $	23.7	$ \\
$\bar{R}^2_{LRw3}	$ & $	5.1	$ & $	21.9	$ & $	33.7	$ & $	36.9	$ & $	1.2	$ & $	5.1	$ & $	11.7	$ & $	22.6	$ & $	2.6	$ & $	14.0	$ & $	31.8	$ & $	35.9	$ & $	0.7	$ & $	3.5	$ & $	11.1	$ & $	23.1	$ \\
$\bar{R}^2_{LRw4}	$ & $	5.0	$ & $	13.5	$ & $	17.3	$ & $	18.5	$ & $	3.2	$ & $	9.1	$ & $	15.5	$ & $	22.3	$ & $	2.4	$ & $	9.0	$ & $	16.5	$ & $	18.1	$ & $	\mathbf{1.7}	$ & $	\mathbf{6.5}	$ & $	14.7	$ & $	22.6	$ \\
$\bar{R}^2_{LRw5}	$ & $	8.0	$ & $	\mathbf{28.4}	$ & $	40.5	$ & $	43.3	$ & $	3.2	$ & $	8.6	$ & $	\mathbf{19.0}	$ & $	\mathbf{32.5}	$ & $	2.5	$ & $	13.8	$ & $	35.4	$ & $	42.5	$ & $	1.2	$ & $	5.2	$ & $	\mathbf{16.1}	$ & $	\mathbf{32.8}	$ \\
\hline																		
\end{tabular}					
\end{center}}																		
\end{table}

\begin{figure}[t]
\centering
\subfigure[Model 1]{\label{F:compara_P4_a} \includegraphics[width=0.4\textwidth]{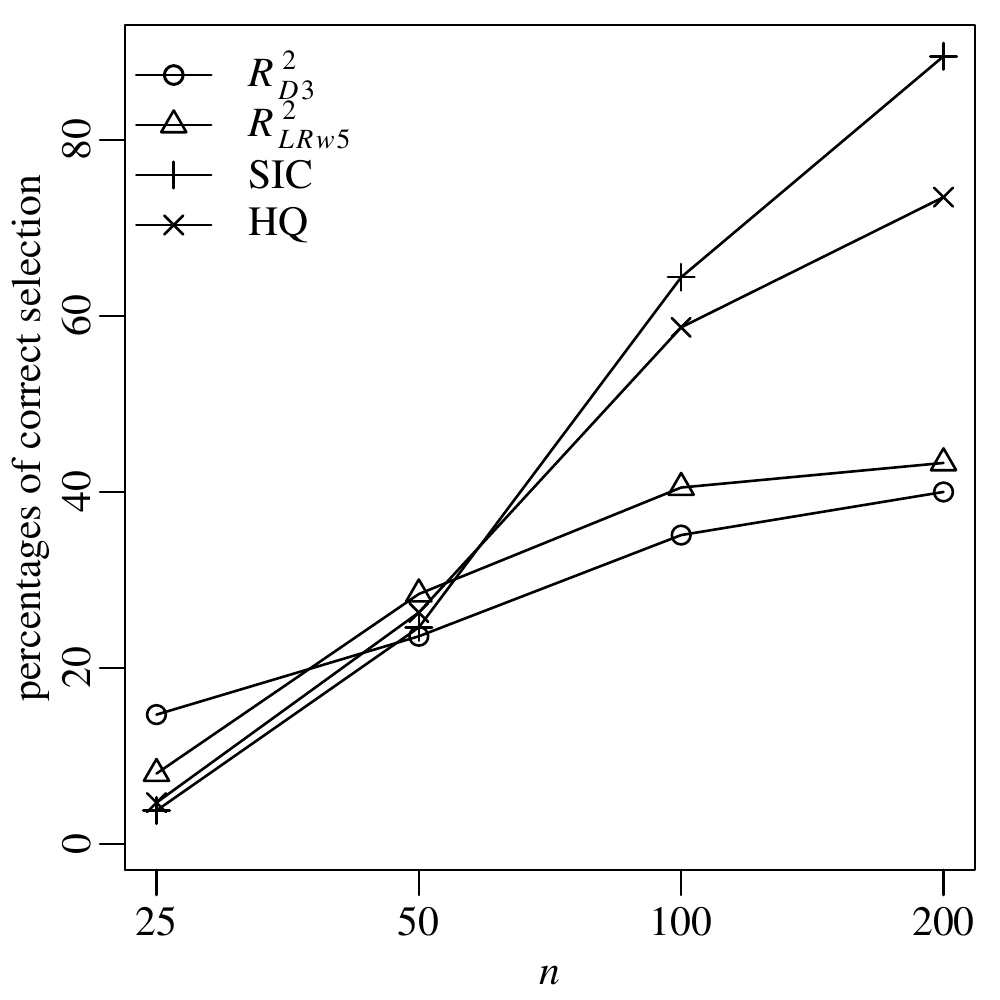}}
\subfigure[Model 2]{\label{F:compara_P4_b} \includegraphics[width=0.4\textwidth]{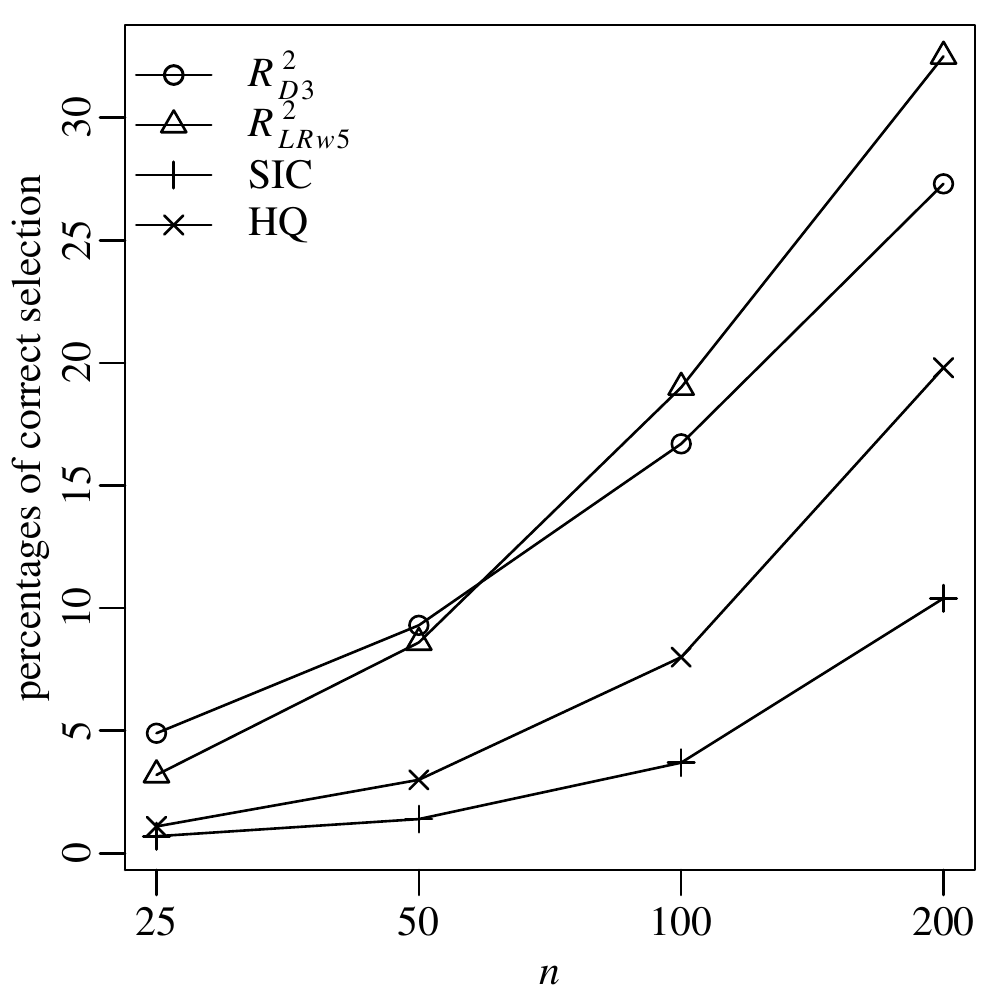}}
\subfigure[Model 3]{\label{F:compara_P4_c} \includegraphics[width=0.4\textwidth]{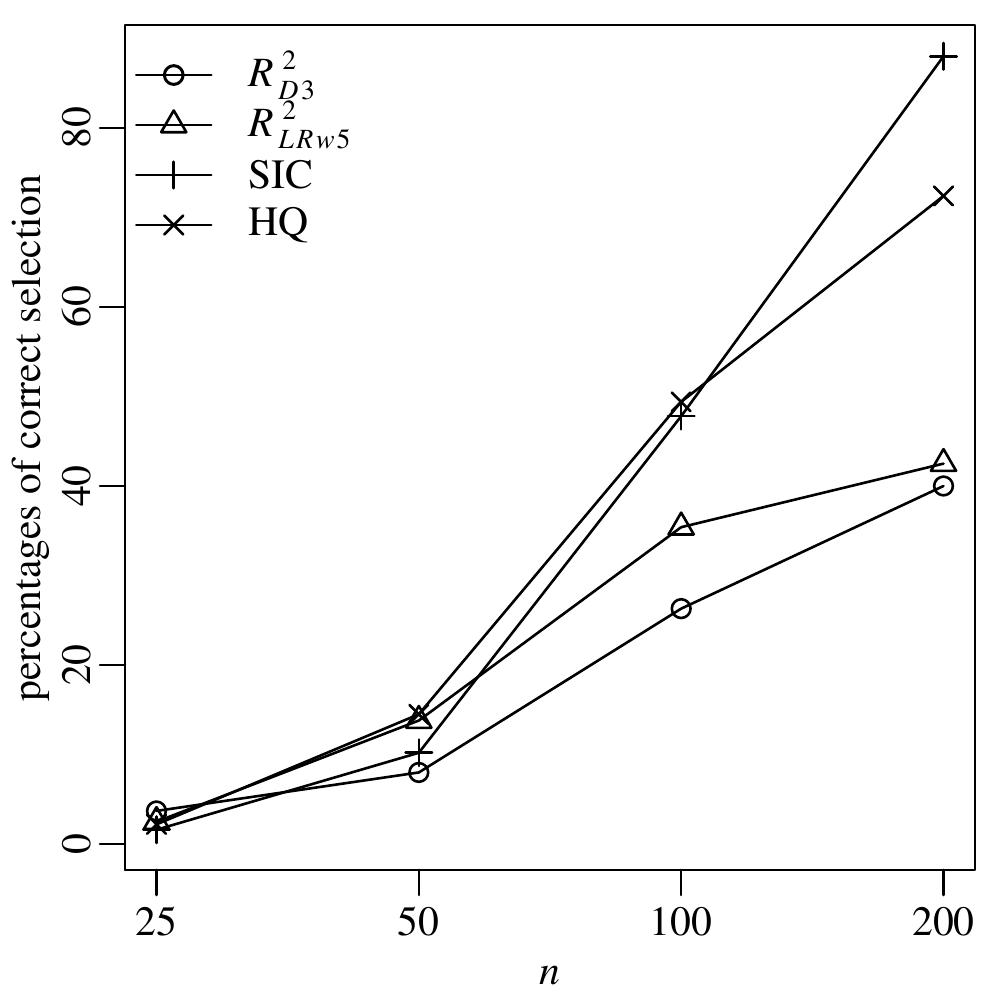}}
\subfigure[Model 4]{\label{F:compara_P4_d} \includegraphics[width=0.4\textwidth]{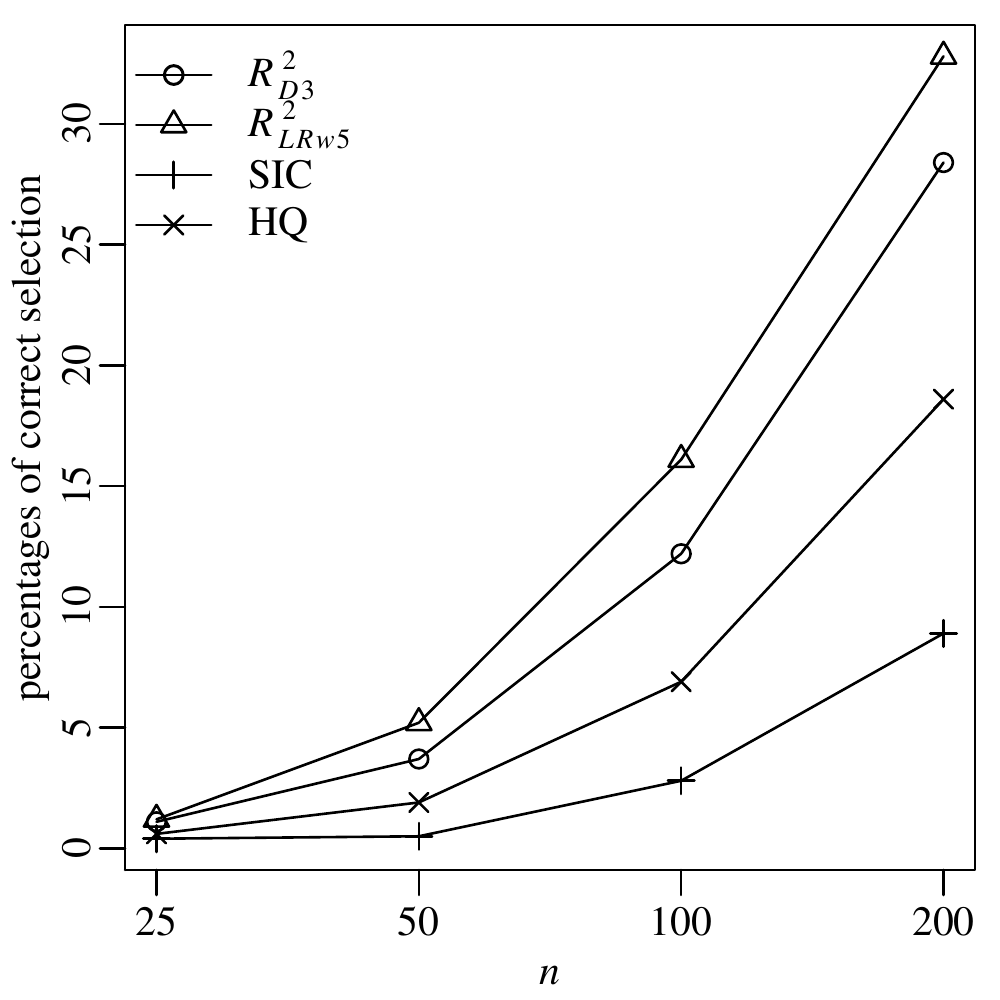}}
\caption{Frequencies (\%) of correct joint model selection: the top four performers.} \label{F:compara_P4}
\end{figure}

\begin{table}[t]
\tablesize																	
{
\caption{Frequencies (\%) of correct dispersion submodel selection when the mean submodel is correctly specified.} \label{T:P3}					
\begin{center}		
\begin{tabular}{l|rrrr|rrrr|rrrr|rrrr}												
\hline																	
 & \multicolumn{4}{c|}{Model $1$}& \multicolumn{4}{c|}{Model $2$}& \multicolumn{4}{c|}{Model $3$}& \multicolumn{4}{c}{Model $4$}\\
\hline	
$n$	& $	25	$ & $	50	$ & $	100	$ & $	200	$ & $	25	$ & $	50	$ & $	100	$ & $	200	$ & $	25	$ & $	50	$ & $	100	$ & $	200	$ & $	25	$ & $	50	$ & $	100	$ & $	200	$ \\
\hline																		
AIC	 & $	10.9	$ & $	38.7	$ & $	64.1	$ & $	70.5	$ & $	4.5	$ & $	7.4	$ & $	17.2	$ & $	34.8	$ & $	11.0	$ & $	38.3	$ & $	64.4	$ & $	70.6	$ & $	4.4	$ & $	7.4	$ & $	16.9	$ & $	34.8	$ \\
AICc	 & $	6.2	$ & $	37.5	$ & $	68.0	$ & $	73.4	$ & $	1.2	$ & $	4.4	$ & $	15.0	$ & $	33.3	$ & $	6.9	$ & $	38.3	$ & $	67.1	$ & $	72.8	$ & $	1.4	$ & $	4.5	$ & $	14.6	$ & $	33.5	$ \\
SIC	 & $	6.9	$ & $	29.4	$ & $	70.5	$ & $	94.5	$ & $	1.7	$ & $	1.9	$ & $	4.1	$ & $	11.0	$ & $	7.6	$ & $	28.8	$ & $	69.0	$ & $	94.0	$ & $	2.0	$ & $	1.8	$ & $	3.8	$ & $	9.5	$ \\
SICc	 & $	1.9	$ & $	22.0	$ & $	67.4	$ & $	\mathbf{95.2}	$ & $	0.2	$ & $	0.7	$ & $	2.4	$ & $	9.2	$ & $	2.0	$ & $	20.9	$ & $	66.4	$ & $	\mathbf{94.6}	$ & $	0.3	$ & $	0.5	$ & $	2.5	$ & $	7.8	$ \\
HQ	 & $	9.8	$ & $	36.2	$ & $	72.4	$ & $	86.1	$ & $	3.5	$ & $	4.2	$ & $	10.1	$ & $	23.3	$ & $	10.0	$ & $	36.7	$ & $	71.1	$ & $	85.6	$ & $	3.6	$ & $	4.2	$ & $	9.6	$ & $	22.3	$ \\
HQc	 & $	4.6	$ & $	32.8	$ & $	\mathbf{73.6}	$ & $	87.7	$ & $	0.8	$ & $	2.0	$ & $	7.5	$ & $	21.3	$ & $	5.0	$ & $	32.2	$ & $	\mathbf{72.2}	$ & $	87.5	$ & $	0.8	$ & $	2.1	$ & $	7.2	$ & $	20.3	$ \\
$ \bar{R}^2_{FC}	$ & $	0.0	$ & $	0.0	$ & $	0.0	$ & $	0.0	$ & $	0.0	$ & $	0.0	$ & $	0.0	$ & $	0.0	$ & $	0.0	$ & $	0.0	$ & $	0.0	$ & $	0.0	$ & $	0.0	$ & $	0.0	$ & $	0.0	$ & $	0.0	$ \\
$\bar{R}^2_{LR}	$ & $	13.1	$ & $	33.7	$ & $	45.1	$ & $	46.9	$ & $	6.9	$ & $	13.2	$ & $	25.1	$ & $	\mathbf{40.4}	$ & $	13.4	$ & $	33.6	$ & $	45.0	$ & $	47.4	$ & $	7.0	$ & $	13.8	$ & $	25.4	$ & $	\mathbf{40.9}	$ \\
$\bar{R}^2_{HS}	$ & $	0.0	$ & $	0.0	$ & $	0.0	$ & $	0.0	$ & $	0.0	$ & $	0.0	$ & $	0.0	$ & $	0.0	$ & $	0.0	$ & $	0.0	$ & $	0.0	$ & $	0.0	$ & $	0.0	$ & $	0.0	$ & $	0.0	$ & $	0.0	$ \\
$\bar{R}^2_{D1}	$ & $	7.0	$ & $	20.4	$ & $	51.4	$ & $	81.3	$ & $	0.4	$ & $	1.0	$ & $	3.3	$ & $	9.6	$ & $	7.2	$ & $	20.1	$ & $	52.0	$ & $	81.5	$ & $	0.6	$ & $	1.3	$ & $	3.1	$ & $	9.5	$ \\
$\bar{R}^2_{D2}	$ & $	7.0	$ & $	19.8	$ & $	51.3	$ & $	81.3	$ & $	0.3	$ & $	0.9	$ & $	3.1	$ & $	9.5	$ & $	7.4	$ & $	19.4	$ & $	51.4	$ & $	81.3	$ & $	0.5	$ & $	1.1	$ & $	3.0	$ & $	9.4	$ \\
$\bar{R}^2_{D3}	$ & $	\mathbf{17.4}	$ & $	26.2	$ & $	38.5	$ & $	43.9	$ & $	6.0	$ & $	10.2	$ & $	18.6	$ & $	30.1	$ & $	\mathbf{17.3}	$ & $	26.1	$ & $	39.0	$ & $	43.7	$ & $	5.1	$ & $	10.7	$ & $	18.8	$ & $	30.6	$ \\
$\bar{R}^2_{D4}	$ & $	16.9	$ & $	26.2	$ & $	38.3	$ & $	43.8	$ & $	6.3	$ & $	10.5	$ & $	18.6	$ & $	30.2	$ & $	17.0	$ & $	26.0	$ & $	38.9	$ & $	43.7	$ & $	5.8	$ & $	10.9	$ & $	18.8	$ & $	30.8	$ \\
$\bar{R}^2_{LR w1}	$ & $	5.1	$ & $	33.3	$ & $	73.0	$ & $	84.4	$ & $	1.0	$ & $	2.4	$ & $	9.3	$ & $	25.4	$ & $	5.7	$ & $	33.2	$ & $	71.7	$ & $	84.0	$ & $	1.1	$ & $	2.5	$ & $	8.7	$ & $	24.4	$ \\
$\bar{R}^2_{LR w2}	$ & $	9.4	$ & $	38.6	$ & $	65.7	$ & $	71.8	$ & $	2.7	$ & $	5.8	$ & $	16.3	$ & $	34.0	$ & $	9.5	$ & $	\mathbf{38.6}	$ & $	65.7	$ & $	71.7	$ & $	2.8	$ & $	6.0	$ & $	15.8	$ & $	34.2	$ \\
$\bar{R}^2_{LR w3}	$ & $	11.4	$ & $	\mathbf{38.8}	$ & $	57.4	$ & $	61.2	$ & $	4.5	$ & $	9.0	$ & $	20.4	$ & $	37.9	$ & $	11.6	$ & $	38.3	$ & $	57.4	$ & $	60.8	$ & $	4.4	$ & $	9.4	$ & $	20.4	$ & $	38.4	$ \\
$\bar{R}^2_{LR w4}	$ & $	11.6	$ & $	25.6	$ & $	31.0	$ & $	32.1	$ & $	\mathbf{10.2}	$ & $	\mathbf{17.1}	$ & $	\mathbf{27.0}	$ & $	38.7	$ & $	12.2	$ & $	24.9	$ & $	30.6	$ & $	32.1	$ & $	\mathbf{10.2}	$ & $	\mathbf{17.5}	$ & $	\mathbf{27.5}	$ & $	39.0	$ \\
$\bar{R}^2_{LR w5}	$ & $	12.4	$ & $	35.9	$ & $	50.4	$ & $	53.3	$ & $	6.0	$ & $	11.2	$ & $	23.2	$ & $	39.6	$ & $	12.5	$ & $	36.1	$ & $	50.7	$ & $	53.0	$ & $	5.9	$ & $	11.7	$ & $	23.1	$ & $	40.0	$ \\

\hline																		
\end{tabular}	
\end{center}}								
\end{table}

\begin{table}
\tablesize																	
{
\caption{Frequencies (\%) of correct mean submodel selection when the dispersion submodel is correctly specified.} \label{T:P2}					
\begin{center}
\begin{tabular}{l|rrrr|rrrr|rrrr|rrrr} 											
\hline																	
 & \multicolumn{4}{c|}{Model $1$}& \multicolumn{4}{c|}{Model $2$}& \multicolumn{4}{c|}{Model $3$}& \multicolumn{4}{c}{Model $4$}\\
\hline
$n$	& $	25	$ & $	50	$ & $	100	$ & $	200	$ & $	25	$ & $	50	$ & $	100	$ & $	200	$ & $	25	$ & $	50	$ & $	100	$ & $	200	$ & $	25	$ & $	50	$ & $	100	$ & $	200	$ \\
\hline
AIC	& $	51.1	$ & $	62.6	$ & $	67.1	$ & $	69.5	$ & $	43.9	$ & $	61.6	$ & $	67.4	$ & $	68.9	$ & $	25.1	$ & $	39.8	$ & $	61.8	$ & $	69.8	$ & $	21.1	$ & $	40.6	$ & $	62.2	$ & $	68.7	$ \\
AICc	& $	73.6	$ & $	73.2	$ & $	72.4	$ & $	71.8	$ & $	68.0	$ & $	74.3	$ & $	73.7	$ & $	71.6	$ & $	23.9	$ & $	\mathbf{41.8}	$ & $	65.3	$ & $	72.2	$ & $	20.6	$ & $	\mathbf{43.1}	$ & $	66.2	$ & $	71.5	$ \\
SIC	& $	67.4	$ & $	85.1	$ & $	91.6	$ & $	95.0	$ & $	58.3	$ & $	83.9	$ & $	91.7	$ & $	94.5	$ & $	23.6	$ & $	36.7	$ & $	70.0	$ & $	93.6	$ & $	21.1	$ & $	37.3	$ & $	72.6	$ & $	94.0	$ \\
SICc	& $	\mathbf{83.4}	$ & $	\mathbf{92.2}	$ & $	\mathbf{94.1}	$ & $	\mathbf{96.1}	$ & $	\mathbf{75.7}	$ & $	\mathbf{92.3}	$ & $	\mathbf{94.6}	$ & $	\mathbf{95.9}	$ & $	14.0	$ & $	30.0	$ & $	67.9	$ & $	\mathbf{94.4}	$ & $	12.5	$ & $	30.3	$ & $	70.8	$ & $	\mathbf{94.8}	$ \\
HQ	& $	56.3	$ & $	73.4	$ & $	81.4	$ & $	85.3	$ & $	48.5	$ & $	72.7	$ & $	81.6	$ & $	85.3	$ & $	25.2	$ & $	40.7	$ & $	69.7	$ & $	85.0	$ & $	21.2	$ & $	42.1	$ & $	71.0	$ & $	85.3	$ \\
HQc	& $	78.6	$ & $	83.1	$ & $	85.7	$ & $	87.4	$ & $	72.2	$ & $	83.7	$ & $	85.9	$ & $	87.4	$ & $	21.1	$ & $	39.2	$ & $	\mathbf{71.8}	$ & $	86.7	$ & $	18.5	$ & $	39.1	$ & $	\mathbf{72.7}	$ & $	86.9	$ \\
$ \bar{R}^2_{FC} $	& $	68.5	$ & $	57.6	$ & $	62.4	$ & $	63.0	$ & $	65.9	$ & $	57.9	$ & $	60.3	$ & $	61.0	$ & $	\mathbf{33.0}	$ & $	38.5	$ & $	54.4	$ & $	60.0	$ & $	\mathbf{31.1}	$ & $	36.6	$ & $	52.9	$ & $	61.4	$ \\
$\bar{R}^2_{LR}$	& $	38.9	$ & $	44.0	$ & $	45.1	$ & $	47.0	$ & $	34.7	$ & $	42.8	$ & $	45.4	$ & $	44.7	$ & $	22.3	$ & $	33.1	$ & $	42.6	$ & $	46.4	$ & $	20.4	$ & $	32.9	$ & $	44.0	$ & $	45.9	$ \\
$\bar{R}^2_{HS}	$ & $	0.7	$ & $	2.1	$ & $	0.0	$ & $	0.0	$ & $	0.0	$ & $	4.0	$ & $	0.0	$ & $	0.0	$ & $	0.2	$ & $	15.2	$ & $	28.7	$ & $	16.5	$ & $	0.0	$ & $	9.0	$ & $	31.5	$ & $	22.2	$ \\
$\bar{R}^2_{D1}$	& $	33.2	$ & $	67.1	$ & $	69.6	$ & $	72.1	$ & $	20.1	$ & $	55.8	$ & $	65.5	$ & $	70.6	$ & $	19.9	$ & $	32.1	$ & $	57.1	$ & $	74.4	$ & $	15.2	$ & $	29.3	$ & $	57.7	$ & $	79.3	$ \\
$\bar{R}^2_{D2}$	& $	66.7	$ & $	89.0	$ & $	89.6	$ & $	90.9	$ & $	47.6	$ & $	84.7	$ & $	86.5	$ & $	89.3	$ & $	23.4	$ & $	32.6	$ & $	67.0	$ & $	91.2	$ & $	18.5	$ & $	29.8	$ & $	65.1	$ & $	92.1	$ \\
$\bar{R}^2_{D3}$		& $	77.7	$ & $	91.1	$ & $	90.9	$ & $	91.7	$ & $	66.7	$ & $	88.7	$ & $	89.0	$ & $	90.5	$ & $	23.5	$ & $	32.3	$ & $	67.5	$ & $	91.7	$ & $	19.0	$ & $	28.8	$ & $	65.5	$ & $	92.6	$ \\
$\bar{R}^2_{D4}$		& $	64.9	$ & $	86.6	$ & $	85.4	$ & $	85.4	$ & $	52.6	$ & $	82.2	$ & $	82.1	$ & $	83.8	$ & $	23.4	$ & $	32.7	$ & $	64.5	$ & $	86.2	$ & $	18.3	$ & $	30.2	$ & $	63.9	$ & $	88.6	$ \\
$\bar{R}^2_{LR w1} $	& $	74.5	$ & $	81.2	$ & $	82.8	$ & $	83.5	$ & $	66.9	$ & $	81.3	$ & $	83.1	$ & $	83.3	$ & $	22.0	$ & $	39.5	$ & $	70.5	$ & $	82.7	$ & $	19.5	$ & $	40.1	$ & $	71.9	$ & $	83.3	$ \\
$\bar{R}^2_{LR w2} $	& $	61.6	$ & $	67.4	$ & $	69.5	$ & $	70.5	$ & $	54.9	$ & $	67.2	$ & $	70.1	$ & $	70.2	$ & $	25.9	$ & $	40.7	$ & $	63.4	$ & $	70.7	$ & $	22.2	$ & $	42.0	$ & $	64.1	$ & $	70.0	$ \\
$\bar{R}^2_{LR w3} $	& $	51.6	$ & $	56.9	$ & $	59.7	$ & $	60.7	$ & $	46.0	$ & $	57.7	$ & $	60.0	$ & $	59.0	$ & $	25.8	$ & $	38.6	$ & $	55.3	$ & $	59.3	$ & $	21.7	$ & $	39.5	$ & $	55.8	$ & $	59.9	$ \\
$\bar{R}^2_{LR w4} $	& $	49.7	$ & $	55.0	$ & $	57.2	$ & $	58.3	$ & $	43.6	$ & $	54.9	$ & $	57.4	$ & $	56.6	$ & $	25.7	$ & $	38.1	$ & $	53.6	$ & $	57.2	$ & $	22.2	$ & $	38.7	$ & $	53.6	$ & $	57.5	$ \\
$\bar{R}^2_{LR w5} $	& $	72.7	$ & $	79.2	$ & $	80.9	$ & $	81.3	$ & $	64.6	$ & $	78.9	$ & $	81.2	$ & $	81.7	$ & $	23.3	$ & $	40.4	$ & $	69.6	$ & $	80.9	$ & $	21.0	$ & $	41.2	$ & $	70.9	$ & $	81.4	$ \\
\hline																		
\end{tabular}																		
\end{center}}								
\end{table}

\begin{table}
\tablesize																		
{
\caption{Frequencies (\%) of correct mean submodel selection when the dispersion was taken to be constant.} \label{T:P5}					
\begin{center}
\begin{tabular}{l|rrrr|rrrr|rrrr|rrrr}												
				
\hline																	
 & \multicolumn{4}{c|}{Model $1$}& \multicolumn{4}{c|}{Model $2$}& \multicolumn{4}{c|}{Model $3$}& \multicolumn{4}{c}{Model $4$}\\
\hline
$n$	& $	25	$ & $	50	$ & $	100	$ & $	200	$ & $	25	$ & $	50	$ & $	100	$ & $	200	$ & $	25	$ & $	50	$ & $	100	$ & $	200	$ & $	25	$ & $	50	$ & $	100	$ & $	200	$ \\	
\hline			
AIC	 & $	70.8	$ & $	68.4	$ & $	73.2	$ & $	74.0	$ & $	66.5	$ & $	66.0	$ & $	70.4	$ & $	72.8	$ & $	32.3	$ & $	40.5	$ & $	63.8	$ & $	71.1	$ & $	30.8	$ & $	37.9	$ & $	61.6	$ & $	73.4	$ \\
AICc	 & $	83.2	$ & $	76.1	$ & $	76.9	$ & $	75.7	$ & $	80.5	$ & $	74.3	$ & $	73.8	$ & $	74.6	$ & $	31.4	$ & $	40.3	$ & $	66.6	$ & $	72.7	$ & $	29.4	$ & $	38.2	$ & $	63.7	$ & $	74.9	$ \\
SIC	 & $	83.2	$ & $	89.4	$ & $	94.6	$ & $	96.7	$ & $	80.9	$ & $	87.3	$ & $	93.6	$ & $	96.9	$ & $	28.3	$ & $	31.8	$ & $	64.4	$ & $	\mathbf{91.7}	$ & $	26.6	$ & $	29.4	$ & $	63.8	$ & $	\mathbf{91.1}	$ \\
SICc	 & $	\mathbf{89.2}	$ & $	93.1	$ & $	\mathbf{95.8}	$ & $	\mathbf{97.2}	$ & $	\mathbf{88.0}	$ & $	92.2	$ & $	\mathbf{95.2}	$ & $	\mathbf{97.3}	$ & $	20.1	$ & $	28.2	$ & $	63.2	$ & $	\mathbf{91.7}	$ & $	19.3	$ & $	25.3	$ & $	62.2	$ & $	91.0	$ \\
HQ	 & $	75.0	$ & $	79.4	$ & $	85.9	$ & $	89.1	$ & $	71.4	$ & $	77.6	$ & $	84.8	$ & $	88.4	$ & $	32.0	$ & $	39.2	$ & $	68.9	$ & $	85.8	$ & $	30.8	$ & $	36.2	$ & $	67.4	$ & $	86.9	$ \\
HQc	 & $	86.1	$ & $	85.9	$ & $	88.7	$ & $	90.1	$ & $	83.8	$ & $	83.5	$ & $	87.5	$ & $	89.5	$ & $	28.5	$ & $	37.2	$ & $	\mathbf{69.2}	$ & $	87.2	$ & $	26.6	$ & $	34.1	$ & $	\mathbf{68.1}	$ & $	87.9	$ \\
$ \bar{R}^2_{FC}	$ & $	54.4	$ & $	45.7	$ & $	50.3	$ & $	49.2	$ & $	49.8	$ & $	44.3	$ & $	48.7	$ & $	49.0	$ & $	30.5	$ & $	32.8	$ & $	45.3	$ & $	48.2	$ & $	29.1	$ & $	31.6	$ & $	43.7	$ & $	50.0	$ \\
$\bar{R}^2_{LR}	$ & $	54.3	$ & $	46.8	$ & $	50.1	$ & $	50.7	$ & $	49.7	$ & $	44.7	$ & $	48.5	$ & $	50.0	$ & $	30.6	$ & $	33.2	$ & $	46.0	$ & $	48.4	$ & $	29.5	$ & $	32.6	$ & $	44.6	$ & $	50.7	$ \\
$\bar{R}^2_{HS}	$ & $	8.3	$ & $	0.6	$ & $	0.0	$ & $	0.0	$ & $	6.1	$ & $	0.2	$ & $	0.0	$ & $	0.0	$ & $	11.6	$ & $	15.4	$ & $	27.6	$ & $	20.4	$ & $	11.2	$ & $	15.1	$ & $	29.5	$ & $	20.6	$ \\
$\bar{R}^2_{D1}	$ & $	85.4	$ & $	92.3	$ & $	91.4	$ & $	90.9	$ & $	84.1	$ & $	90.4	$ & $	89.3	$ & $	91.1	$ & $	20.2	$ & $	27.5	$ & $	61.0	$ & $	87.9	$ & $	19.1	$ & $	24.2	$ & $	57.8	$ & $	86.3	$ \\
$\bar{R}^2_{D2}	$ & $	89.0	$ & $	\mathbf{94.6}	$ & $	95.3	$ & $	96.2	$ & $	87.5	$ & $	93.4	$ & $	94.5	$ & $	96.2	$ & $	18.4	$ & $	26.0	$ & $	59.8	$ & $	90.1	$ & $	18.5	$ & $	23.7	$ & $	59.2	$ & $	89.4	$ \\
$\bar{R}^2_{D3}	$ & $	\mathbf{89.2}	$ & $	\mathbf{94.6}	$ & $	95.4	$ & $	96.2	$ & $	87.6	$ & $	\mathbf{93.5}	$ & $	94.5	$ & $	96.2	$ & $	18.4	$ & $	26.0	$ & $	59.7	$ & $	90.2	$ & $	18.4	$ & $	23.5	$ & $	59.1	$ & $	89.5	$ \\
$\bar{R}^2_{D4}	$ & $	88.3	$ & $	94.0	$ & $	94.0	$ & $	94.4	$ & $	86.8	$ & $	92.7	$ & $	92.9	$ & $	94.8	$ & $	18.7	$ & $	26.7	$ & $	60.6	$ & $	89.5	$ & $	18.5	$ & $	23.6	$ & $	58.8	$ & $	88.7	$ \\
$\bar{R}^2_{LRw1}	$ & $	85.8	$ & $	85.3	$ & $	86.7	$ & $	87.1	$ & $	83.3	$ & $	82.9	$ & $	85.4	$ & $	86.2	$ & $	27.0	$ & $	37.2	$ & $	69.0	$ & $	84.0	$ & $	25.6	$ & $	34.0	$ & $	67.7	$ & $	85.0	$ \\
$\bar{R}^2_{LRw2}	$ & $	76.6	$ & $	71.9	$ & $	74.7	$ & $	74.9	$ & $	72.9	$ & $	70.2	$ & $	72.0	$ & $	73.6	$ & $	32.8	$ & $	\mathbf{40.6}	$ & $	65.2	$ & $	71.8	$ & $	31.2	$ & $	\mathbf{38.2}	$ & $	62.6	$ & $	74.4	$ \\
$\bar{R}^2_{LRw3}	$ & $	68.1	$ & $	61.4	$ & $	64.6	$ & $	64.7	$ & $	63.7	$ & $	59.4	$ & $	62.6	$ & $	63.6	$ & $	33.3	$ & $	39.5	$ & $	57.9	$ & $	62.0	$ & $	31.5	$ & $	37.1	$ & $	55.8	$ & $	64.2	$ \\
$\bar{R}^2_{LRw4}	$ & $	67.5	$ & $	59.8	$ & $	62.6	$ & $	62.7	$ & $	63.1	$ & $	57.7	$ & $	60.3	$ & $	61.5	$ & $	\mathbf{33.6}	$ & $	39.0	$ & $	56.3	$ & $	60.1	$ & $	\mathbf{31.8}	$ & $	36.9	$ & $	54.3	$ & $	61.7	$ \\
$\bar{R}^2_{LRw5}	$ & $	85.5	$ & $	84.1	$ & $	84.9	$ & $	85.6	$ & $	83.3	$ & $	81.6	$ & $	84.0	$ & $	84.3	$ & $	28.0	$ & $	38.3	$ & $	69.2	$ & $	82.2	$ & $	26.4	$ & $	35.2	$ & $	67.6	$ & $	83.5	$ \\

\hline
\end{tabular}
\end{center}}								
\end{table}

The figures in Table~\ref{T:P4} show that joint selection of the regressors in both submodels is typically not accurate when the sample size is small and/or the dispersion submodel is weakly identifiable. Notice, for instance, the small frequency of correct model selection when $n=25$, especially in Models 2 and 4. Model selection based on $\bar{R}^2_{D3}$ works well in small sample in all models. In Models 1 and 3 (dispersion submodel is easily identifiable) the SIC achieves reliable model selection. $\bar{R}^2_{LRw5}$ also delivers accurate model selection in some situations. The criteria $\bar{R}^2_{D3}$ and $\bar{R}^2_{LRw5}$ display good performance under weak identifiability of the mean submodel and in small samples; however, their performances are poor otherwise. Model selection via the SIC is accurate when in large samples and when the mean submodel is easily identifiable; otherwise, it does not perform well. Overall, the best performer is the HQ criterion. It delivers reliable model selection in nearly all scenarios, thus having a well balanced performance. 

Figure~\ref{F:compara_P4} displays the frequencies of correct model selection achieved by the $\bar{R}^2_{D3}$, $\bar{R}^2_{LRw5}$, SIC and HQ criteria. The top performers when the dispersion submodel is weakly identifiable are $\bar{R}^2_{D3}$ and $\bar{R}^2_{LRw5}$. 
In Models 1 and 3 and when $n=100,200$, the SIC delivers the most accurate model selection, being closely followed by HQ. We note that HQ also displays good performance in Models 2 and 4. We thus recommend that joint selection of the regressors in the two submodels be based on $\bar{R}^2_{D3}$ or $\bar{R}^2_{LRw5}$ when $n \leq 50$ and on HQ for larger samples.

Notice that the finite sample performances of the different model selection criteria for the joint selection of the regressors of both submodels are heavily dependent on the identifiability of such submodels. It is also noteworthy that the best joint model selection strategies are not necessarily the most accurate when model selection focuses on one of the submodels; see the results in Tables~\ref{T:P3} and \ref{T:P2}.

Table~\ref{T:P3} presents the simulation results obtained when the mean submodel is correctly specified and we  focus on the dispersion model selection. The best performer in Models 2 and 4 (dispersion submodel is weakly identifiable) is $\bar{R}^2_{LR w4}$. When Models 1 and 3 (dispersion submodel is easily identifiable) are used as data generating processes, the best performer is $\bar{R}^2_{LR w3}$ when $n=50$; when $n=100$, the best performer is the HQc; when $n=200$, the winner is the SICc. The figures in Table~\ref{T:P3} show that the AIC delivers reliable model selection since it is always among the best performers. 

We now move to the situation in which the dispersion submodel is correctly specified and model selection takes place in the mean submodel. The corresponding numerical results are presented in Table~\ref{T:P2}. The best performer when the mean submodel is easily identifiable (Models 1 and 2) is the SICc. When the mean submodel is weakly identifiable (Models 3 and 4) the most accurate model selection was achieved using $\bar{R}^2_{FC}$ when $n=25$, AICc when $n=50$, HQc when $n=100$, and SICc when $n=200$. We note that $\bar{R}^2_{D3}$ performed well in all scenarios. Overall, the most reliable criteria here are $\bar{R}^2_{D3}$, SICc and HQc. 

Our next set of Monte Carlo results were obtained by taking dispersion to be constant and focusing on selecting covariates for the mean submodel. The results are presented in Table~\ref{T:P5}. 
It is interesting to note that in some cases mean submodel selection is more accurate when dispersion is taken to be constant than when the dispersion submodel is correctly specified, especially when the sample size is small ($n=25,50$) and the model is easily identifiable (Models 1 and 2). Compare, for instance, the frequencies of correct model selection for Model 2 with $n=25$ in Tables~\ref{T:P2} and \ref{T:P5}. 
The SIC frequency of correct model selection when dispersion is taken to be constant is nearly 15\% larger than when the dispersion submodel is correctly identified ($88.0\%$ vs.\ $75.7\%$). Overall, the frontrunners are the SICc, $\bar{R}^2_{D3}$ and $\bar{R}^2_{LR w2}$. We also note that the HQc performed well in several scenarios. 

The results presented so far indicate that the best performing model selection criteria for selecting regressors for the mean and dispersion submodels do not typically coincide. That is, the best modeling strategies for the mean submodel are not the best ones when it comes to selecting covariates for the dispersion submodel. This fact may explain the poor performances of the different model selection criteria when used to jointly select regressors for both submodels; see Table~\ref{T:P4}. It is also noteworthy that some of the criteria perform quite well when one takes dispersion to be constant and focuses on selecting covariates for the mean submodel. 
Based on such evidence, 
the proposed fast two step model selection procedure, presented in Section~\ref{ss:two-step}, arises naturally. 

We shall now present Monte Carlo evidence on the proposed model selection scheme. We used different combinations of criteria in Steps (1) and (2), based on the numerical evidence already presented. The following implementations of the proposed scheme (PS) were considered: 
\begin{enumerate}[]
\item $\text{PS}_1$: SICc is used in Step (1) and $\bar{R}^2_{LR w4}$ is used in Step (2);

\item $\text{PS}_2$: SICc is used in Step (1) and $\bar{R}^2_{D3}$ is used in Step (2);

\item $\text{PS}_3$: SICc is used in Step (1) and SICc is used in Step (2);

\item $\text{PS}_4$: HQc is used in Step (1) and HQc is used in Step (2);

\item $\text{PS}_5$: AIC is used in Step (1) and $\bar{R}^2_{LR w4}$ is used in Step (2);

\item $\text{PS}_6$: $\bar{R}^2_{LR w4}$ is used in Step (1) and $\bar{R}^2_{D3}$ is used in Step (2);

\item $\text{PS}_7$: $\bar{R}^2_{LR w5}$ is used in Step (1) and $\bar{R}^2_{LR w5}$ is used in Step (2).
\end{enumerate}

\begin{table}[t]
\tablesize																		
{
\caption{Frequencies (\%) of correct model selected using the proposed two step scheme.} \label{T:steps}	
\begin{center}	
\begin{tabular}{l|rrrr|rrrr|rrrr|rrrr}
\hline																	
 & \multicolumn{4}{c|}{Model $1$}& \multicolumn{4}{c|}{Model $2$}& \multicolumn{4}{c|}{Model $3$}& \multicolumn{4}{c}{Model $4$}\\
\hline
$n$	& $	25	$ & $	50	$ & $	100	$ & $	200	$ & $	25	$ & $	50	$ & $	100	$ & $	200	$ & $	25	$ & $	50	$ & $	100	$ & $	200	$ & $	25	$ & $	50	$ & $	100	$ & $	200	$ \\	
\hline			
$\text{PS}_1$ & $	10.5	$ & $	24.2	$ & $	29.6	$ & $	31.2	$ & $	\mathbf{9.0}	$ & $	\mathbf{15.6}	$ & $	\mathbf{25.7}	$ & $	\mathbf{37.7}	$ & $	2.4	$ & $	7.0	$ & $	19.7	$ & $	29.5	$ & $	1.7	$ & $	5.2	$ & $	\mathbf{16.6}	$ & $	\mathbf{35.5}	$ \\
$\text{PS}_2$ & $	\mathbf{15.8}	$ & $	24.6	$ & $	36.7	$ & $	42.5	$ & $	5.2	$ & $	9.5	$ & $	17.7	$ & $	29.5	$ & $	2.8	$ & $	6.3	$ & $	24.2	$ & $	40.2	$ & $	1.0	$ & $	2.9	$ & $	11.5	$ & $	27.9	$ \\
$\text{PS}_3$ & $	1.6	$ & $	20.7	$ & $	64.6	$ & $	\mathbf{92.5}	$ & $	0.2	$ & $	0.6	$ & $	2.3	$ & $	8.9	$ & $	0.2	$ & $	5.0	$ & $	40.8	$ & $	\mathbf{86.8}	$ & $	0.0	$ & $	0.1	$ & $	1.6	$ & $	7.1	$ \\
$\text{PS}_4$ & $	3.8	$ & $	28.4	$ & $	\mathbf{65.4}	$ & $	79.1	$ & $	0.6	$ & $	1.6	$ & $	6.6	$ & $	18.9	$ & $	1.1	$ & $	11.4	$ & $	\mathbf{49.9}	$ & $	76.2	$ & $	0.2	$ & $	0.7	$ & $	5.1	$ & $	17.6	$ \\
$\text{PS}_5$ & $	8.5	$ & $	17.6	$ & $	22.5	$ & $	24.0	$ & $	6.7	$ & $	11.4	$ & $	19.1	$ & $	28.6	$ & $	4.0	$ & $	10.2	$ & $	20.1	$ & $	22.7	$ & $	\mathbf{3.2}	$ & $	\mathbf{7.1}	$ & $	\mathbf{16.6}	$ & $	28.8	$ \\
$\text{PS}_6$ & $	12.8	$ & $	15.9	$ & $	23.6	$ & $	27.6	$ & $	3.9	$ & $	6.2	$ & $	11.4	$ & $	19.3	$ & $	\mathbf{5.3}	$ & $	9.7	$ & $	21.8	$ & $	26.0	$ & $	1.4	$ & $	4.2	$ & $	10.5	$ & $	19.2	$ \\
$\text{PS}_7$ & $	10.9	$ & $	\mathbf{30.9}	$ & $	42.9	$ & $	45.6	$ & $	4.9	$ & $	9.3	$ & $	19.6	$ & $	33.4	$ & $	3.4	$ & $	\mathbf{13.5}	$ & $	35.3	$ & $	43.2	$ & $	1.4	$ & $	4.0	$ & $	15.8	$ & $	33.5	$ \\
\hline																	
\end{tabular}																	
\end{center}}								
\end{table}

\begin{figure}[t]
\centering
\subfigure[Model 1]{\label{F:compara_P4_P10_a} \includegraphics[width=0.4\textwidth]{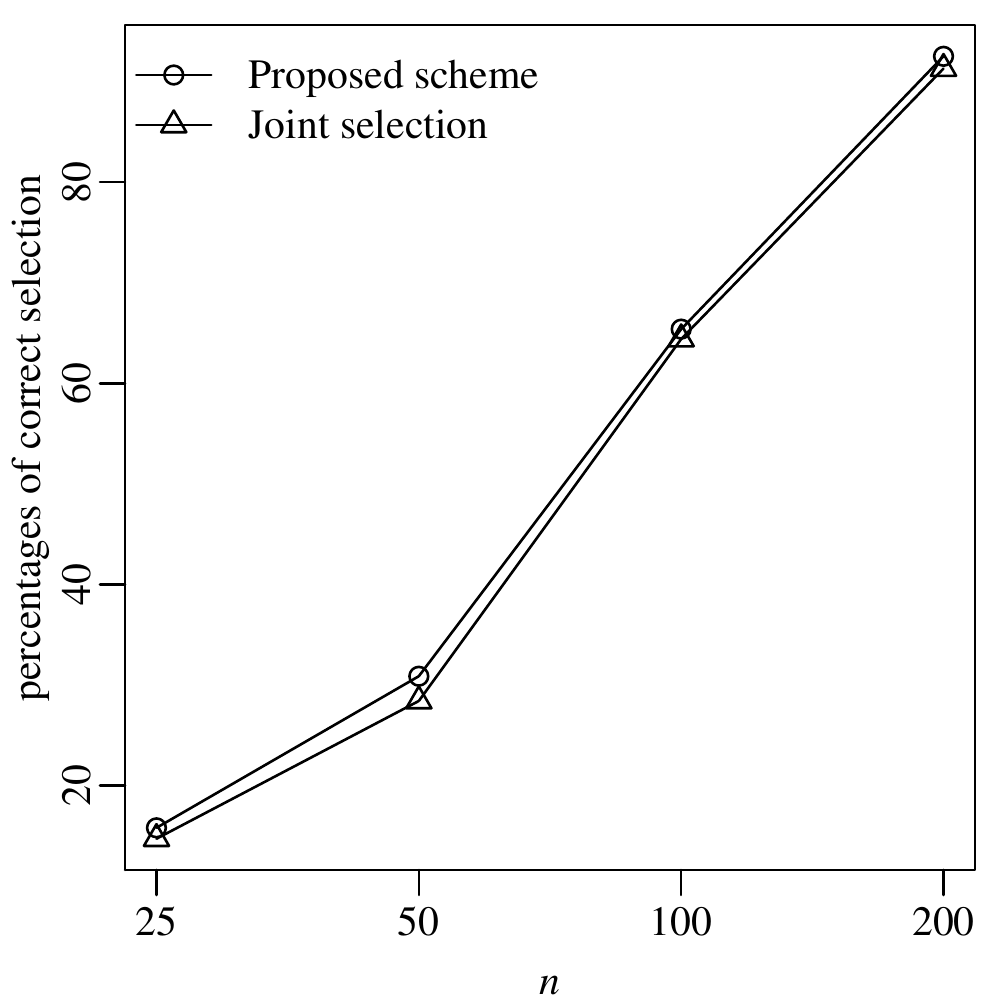}}
\subfigure[Model 2]{\label{F:compara_P4_P10_b} \includegraphics[width=0.4\textwidth]{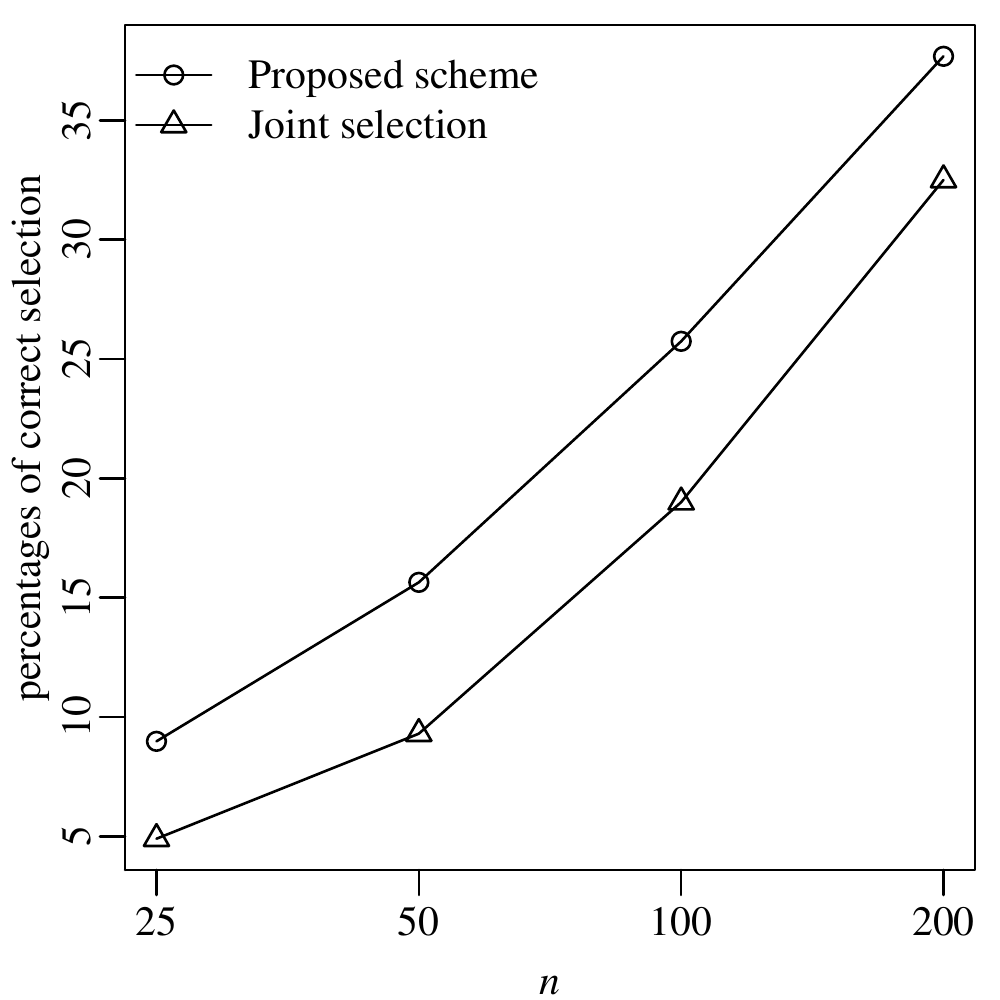}}
\subfigure[Model 3]{\label{F:compara_P4_P10_c} \includegraphics[width=0.4\textwidth]{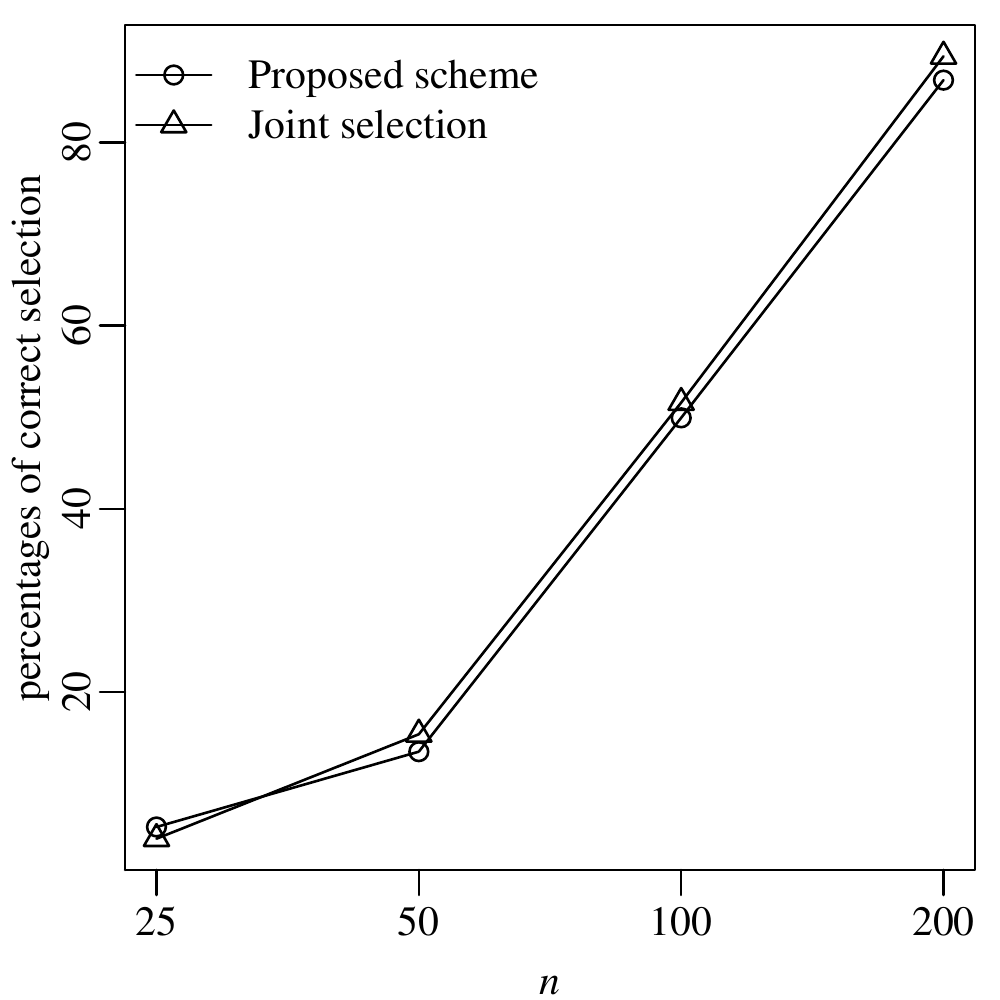}}
\subfigure[Model 4]{\label{F:compara_P4_P10_d} \includegraphics[width=0.4\textwidth]{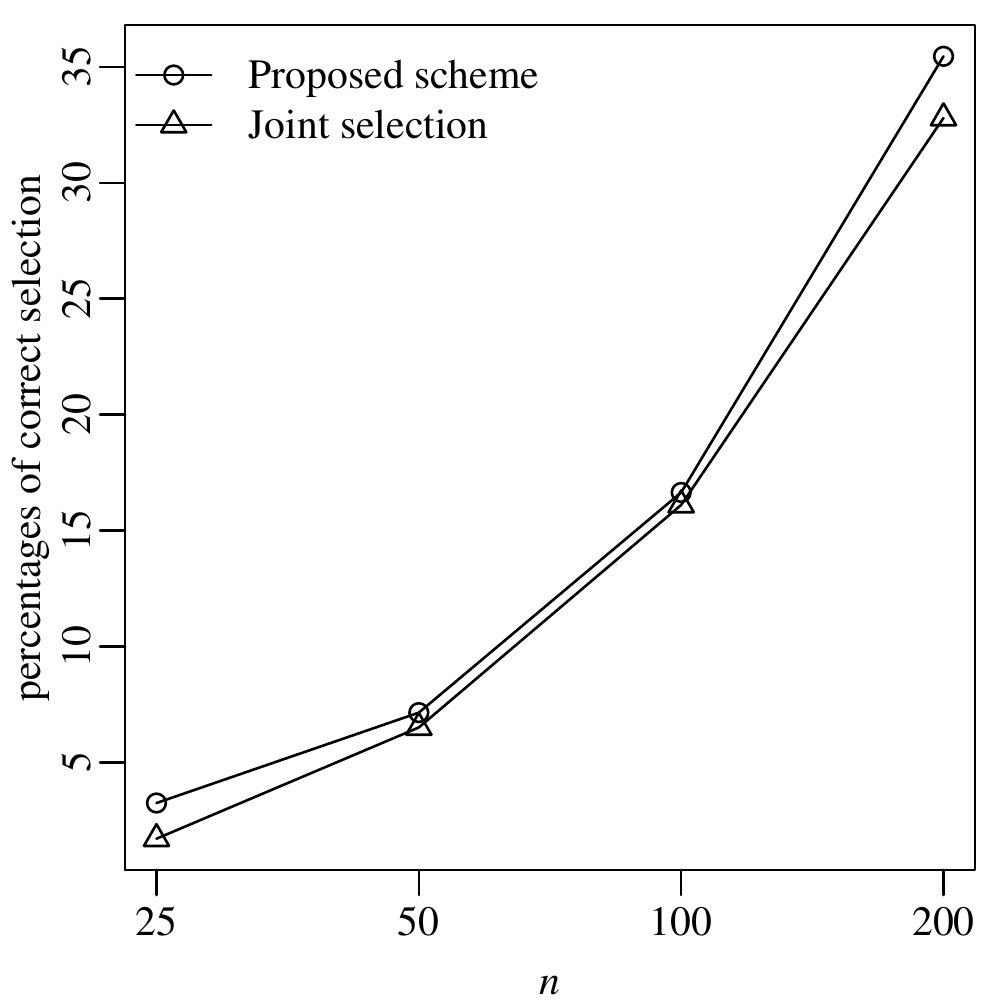}}
\vspace{-0.2cm}
\caption{Frequencies (\%) of correct model selection: proposed two step and joint model selection.}\label{F:compara_P4_P10}
\end{figure}

Monte Carlo results are presented in Table~\ref{T:steps}. By comparing these results to those reported in Tables~\ref{T:steps} and \ref{T:P4} we note that the proposed model selection scheme is more accurate in nearly all scenarios. The frequencies of correct model selection of the two best performers in each case (proposed scheme and joint model selection) are displayed in Figure~\ref{F:compara_P4_P10}. 
Among all considered implementations of the PS, the scheme $\text{PS}_1$ is the best performer when the dispersion submodel is weakly identifiable (Models 2 and 4). When it is easily identifiable (Models 1 and 3), the most accurate model selection scheme is: $\text{PS}_6$ for $n=25$, $\text{PS}_7$ for $n=50$, $\text{PS}_3$ for $n=100$ and $\text{PS}_4$ for $n=200$.

\subsection{Final discussion and guideline for choose model selection criteria}\label{guide}

As a final remark, we emphasize that correct specification of the dispersion submodel is the most critical step in varying dispersion beta regression model selection. 
Notice, for instance, that the frequencies of correct model selection are considerably lower in Models 2 and  4 (dispersion submodel weakly identifiable) than in Models 1 and 3 (dispersion submodel easily identifiable); see Table~\ref{T:steps}. 
The identifiability of the model and the sample size directly influence in performances of the model selection criteria. 

The proposed two step model selection scheme is computationally more efficient than the usual approach
and performs equally well or even better. Additionally, based on our numerical results, we suggest the use of the following criterion: 
\begin{enumerate}
\item In small samples ($n \leq 50$): use $\text{PS}_1$ or $\text{PS}_5$;
\item In large samples ($n > 50$): use $\text{PS}_4$.
\end{enumerate}

In addition to using our model selection scheme, we recommend that practitioners check whether the selected model is correctly specified. To that end, we recommend that they use the misspecification test introduced by~\cite{Pereira2013}.

\section{An empirical application}\label{S:application}

In what follows we shall present the results of an empirical application. We use data from a study of reading ability in a group of $44$ Australian children that attended primary school \citep{Pammer2004}. These data were also analyzed by \cite{Smithson2006} and \cite{Ferrari2011}. 
The response ($y$) are reading accuracy indices of such children. The independent variables are: nonverbal IQ converted to $z$-scores ($x_2$) and dyslexia versus non-dyslexia status ($x_3$). The participants (19 dyslexics and 25 controls) were students from primary schools in the Australian Capital Territory. Their ages range from eight years five months to twelve years three months. The covariate $x_3$ is a dummy variable, which equals 1 if the child is dyslexic and $-1$ otherwise. 
As in \cite{Smithson2006} and \cite{Ferrari2011}, 
the 
observed scores were linearly transformed from their original scale to the open unit interval $(0,1)$. 
Computer code for two-step model selection and the data used in this application are available at \url{http://www.ufsm.br/bayer/auto-beta-reg.zip}.

In \cite{Smithson2006}, the authors consider a third covariate ($x_4$), namely: the interaction between $x_2$ and $x_3$, that is, $x_4 = x_2 \times x_3$. At the outset, the authors estimate linear regression models and then estimate a fixed dispersion beta regressions. However, they conclude that the inferential results may be inaccurate given that dispersion is not constant. They then estimate a varying dispersion beta regression model.

We consider a varying dispersion beta regression model with logit links in the two submodels. In addition to the covariates described above, we also consider $x_5=x_2^2$ and $x_6=x_3 \times x_5$. Since there are five candidate covariates, we need to consider $2\times(2^5 +1)=66$ models in the model selection procedure proposed in this paper and $(2^5 +1)^2=1089$ candidate models when carrying out joint model selection. 

We start by testing the null hypothesis of constant dispersion using a score test; see Section~\ref{S:model_beta_var} for details on such a test. The mean submodel includes the following covariates: $x_2$, $x_3$ and $x_4$. The null hypothesis under test is $\mathcal{H}_0 : \gamma_2=\gamma_3=\gamma_4= 0$, where ${\rm logit}(\sigma_t)=\gamma_1+\gamma_2 x_2+\gamma_3 x_3+\gamma_4 x_4$. The score test statistic equals  18.069, the test $p$-value being 0.0004. We thus reject the null hypothesis of constant dispersion at the usual nominal levels. 

Notice that the sample size is close to 50 and that our numerical evidence indicates that for this sample size the best performing model two step selection schemes are $\text{PS}_1$ and $\text{PS}_5$. When the $\text{PS}_1$ scheme is used we arrive at a model that only includes one covariate in the mean and dispersion submodels, namely: $x_3$. Standard diagnostic analysis, however, indicates that the model is not correctly specified. Using $\text{PS}_5$, with AIC in step (1) and $\bar{R}^2_{LR w4}$ in step (2), we arrive at a beta regression model that uses $x_3$, $x_5$ and $x_6$ as mean covariates and $x_2$, $x_3$ $x_4$ and $x_5$ as dispersion covariates. All covariates are statistically significant at the usual nominal levels; see Table~\ref{T:ajuste_aplica}.

\begin{table}
\footnotesize																		
{

\caption{Parameter estimates of the beta regression model with varying dispersion; reading ability data.} \label{T:ajuste_aplica}
\begin{center}																				
\begin{tabular}{lrrrr}																		
\hline
Parameter	& 	Estimate	 & 	Std. error &	$z$ stat & $p$-value  \\
\hline																						
\multicolumn{5}{c}{Submodel of $\mu$}\\
\hline
$\beta_1$ (Constant) & $1.0494$ & $0.1605$ & $6.539$ & $0.0000$ \\
$\beta_3$ (Dyslexia) & $-0.8587$ & $0.1587$ & $-5.411$ & $0.0000$ \\
$\beta_5$ (${\rm IQ}^2$)& $0.4524$ & $0.0580$ & $7.804$ & $0.0000$ \\
$\beta_6$ (Dyslexia$\, \times {\rm IQ}^2$)& $-0.3866$ & $0.0576$ & $-6.720$ & $0.0000$ \\
\hline																						
\multicolumn{5}{c}{Submodel of $\sigma$}\\
\hline 
$\gamma_1$ (Constant) & $-1.0072$ & $0.1828$ & $-5.509$ & $0.0000$ \\
$\gamma_2$ (IQ) & $-0.9259$ & $0.1498$ & $-6.180$ & $0.0000$ \\
$\gamma_3$ (Dyslexia) & $-0.9047$ & $0.1603$ & $-5.645$ & $0.0000$ \\
$\gamma_4$ (Dyslexia$\, \times \,$QI) & $-0.8559$ & $0.2633$ & $-3.251$ & $0.0025$ \\
$\gamma_5$ (${\rm IQ}^2$)& $-1.1005$ & $0.2065$ & $-5.328$ & $0.0000$ \\
\hline 
\multicolumn{5}{c}{$R^2_{FC}=0.63$} \\
\multicolumn{5}{c}{$R^2_{LR}=0.88$}\\
\hline																		
\end{tabular}			
\end{center}}			
\end{table}

It is noteworthy that $R^2_{FC}$ and $R^2_{LR}$ differ considerably: $R^2_{FC} = 0.63$ and $R^2_{LR} = 0.88$. This happens because $R^2_{FC}$ is less sensitive to the dispersion model specification, unlike $R^2_{LR}$, which assumes significantly larger values when the dispersion submodel is correctly selected. The two measures tend to assume similar values in constant dispersion beta regressions. We recommend the use of $R^2_{LR}$ in varying dispersion models. 

The beta regression model whose parameter estimates are presented in Table~\ref{T:ajuste_aplica} differs from the model used in \cite{Smithson2006}. The authors model the precision parameter $\phi$ (and not the dispersion parameter $\sigma$) using as link function $-{\rm ln}(\cdot)$. Their mean submodel uses as regressors $x_2$, $x_3$  and $x_4$ and their precision submodel includes $x_2$ and $x_3$ as covariates. 
Indeed, 
these are the same covariates for the selected model using the two step scheme considering only $x_2$, $x_3$ and $x_4$ as candidate covariates. 
However, the diagnostic analysis of this model, as shown in \cite{CribariQueiroz2014}, evidences some problems and 
its $R^2_{FC}$ and $R^2_{LR}$ measures are considerably smaller than those of our selected model in Table~\ref{T:ajuste_aplica}. 
In \cite{CribariQueiroz2014}, bootstrap-based testing inferences also suggested that ${\rm IQ}^2$ must be included in the model.

\section{Conclusions}\label{S:conclusions}

This paper addressed the issue of model selection in varying dispersion beta regressions. We presented several model selection criteria that can be used in beta regression modeling and proposed two new model selection criteria that explicitly account for varying dispersion. We also proposed a fast two step model selection procedure that outperforms joint model selection, i.e., the joint selection of the covariates that must enter the mean and dispersion submodels. The proposed model selection scheme is also much less costly from a computational viewpoint than the joint model selection. We have also presented the results of extensive Monte Carlo simulations and guidelines for choosing a model selection criteria in Section~\ref{guide}. The results show that the finite sample performances of the different model selection approaches are typically strongly dependent on the model identifiability. We also argue that it is more appropriate to use $R^2_{LR}$ as a pseudo-$R^2$ measure in varying dispersion beta regressions than $R^2_{FC}$ since the former is more sensitive to the specification of the dispersion submodel. Finally, we an empirical application was performed.

\section*{Acknowledgements}

We gratefully acknowledge partial financial support from CAPES, CNPq, and FAPERGS, Brazil. We also thank two anonymous referees for their comments and suggestions.

\singlespacing

\bibliographystyle{arxiv}

\bibliography{betareg}

\end{document}